\documentclass[sigconf]{acmart}

\usepackage{tikz}
\usepackage{bm}
\usepackage{rotating}

\usepackage{amsmath,amsfonts,latexsym, amssymb}
\usepackage{amsthm}
\usepackage{algorithmic}
\usepackage{graphicx}
\usepackage{textcomp}
\usepackage{xcolor}
\usepackage{multirow}
\usepackage{hyperref}
\usepackage{hyperxmp}
\usepackage{url}
\usepackage{float}
\usepackage{relsize}
\usepackage{rotating}
\usepackage[switch]{lineno}
\usepackage{booktabs}
\usepackage{appendix}
\usepackage{xcolor}
\usepackage{ulem}

\newtheorem{theorem}{Theorem}[section]

\newtheorem{lemma}[theorem]{Lemma}
\newtheorem{mydef}{Definition}

\DeclareMathOperator{\Tr}{Tr}

\DeclareMathOperator{\T}{\bm{\Theta}}

\DeclareMathOperator{\B}{\mathcal{B}}
\DeclareMathOperator{\BB}{\mathbf{B}}
\DeclareMathOperator{\bb}{\mathbf{b}}

\DeclareMathOperator{\J}{\mathbf{J}}

\DeclareMathOperator{\La}{\bm{\Lambda}}

\AtBeginDocument{%
  }

\setcopyright{acmlicensed}
\copyrightyear{2018}
\acmYear{2018}
\acmDOI{XXXXXXX.XXXXXXX}
\acmConference[Conference acronym 'XX]{Make sure to enter the correct
  conference title from your rights confirmation email}{June 03--05,
  2018}{Woodstock, NY}
\acmISBN{978-1-4503-XXXX-X/2018/06}

\renewcommand\footnotetextcopyrightpermission[1]{}
\begin{document}

\title{PROVGEN: A Privacy-Preserving Approach for Outcome Validation in Genomic Research}


\author{Yuzhou Jiang}
\email{yxj466@case.edu}
\affiliation{%
  \institution{Case Western Reserve University}
  \city{Cleveland}
  \state{Ohio}
  \country{USA}
}
\author{Tianxi Ji}
\email{tiji@ttu.edu}
\affiliation{%
  \institution{Texas Tech University}
  \city{Lubbock}
  \state{Texas}
  \country{USA}
}
\author{Erman Ayday}
\email{exa208@case.edu}
\affiliation{%
  \institution{Case Western Reserve University}
  \city{Cleveland}
  \state{Ohio}
  \country{USA}
}

\renewcommand{\shortauthors}{Jiang et al.}

\begin{abstract}
As genomic research has grown increasingly popular in recent years, dataset sharing has remained limited due to privacy concerns. This limitation hinders the reproducibility and validation of research outcomes, both of which are essential for identifying computational errors during the research process.
In this paper, we introduce PROVGEN, a privacy-preserving method for sharing genomic datasets that facilitates reproducibility and outcome validation in genome-wide association studies (GWAS).
Our approach encodes genomic data into binary space and applies a two-stage process. First, we generate a differentially private version of the dataset using an XOR-based mechanism that incorporates biological characteristics. Second, we restore data utility by adjusting the Minor Allele Frequency (MAF) values in the noisy dataset to align with published MAFs using optimal transport. Finally, we convert the processed binary data back into its genomic representation and publish the resulting dataset. We evaluate PROVGEN on three real-world genomic datasets and compare it with local differential privacy and three synthesis-based methods. We show that our proposed scheme outperforms all existing methods in detecting GWAS outcome errors, achieves better data utility, and provides higher privacy protection against membership inference attacks (MIAs). By adopting our method, genomic researchers will be inclined to share differentially private datasets while maintaining high data quality for reproducibility of their findings.
\end{abstract}


\maketitle

\section{Introduction}
\label{sec:intro}
Recent advancements in genome sequencing have unlocked significant research opportunities in genomics. Through computational and statistical methods, such as genome-wide association studies (GWAS), researchers have identified numerous associations between diseases/traits and genes, further enriching our understanding of the field.

As genomic research becomes increasingly popular, unintentional errors may occur during the research process or when reporting outcomes. Thus, medical professionals, who often rely on GWAS outcomes for clinical applications such as treatment procedures, need to ensure that these results are accurately computed. However, validating these results is challenging, as they typically lack access to the original datasets due to privacy concerns regarding genomic datasets~\cite{lin2004genomic}. As a result, the inability to validate the outcomes compromises the assessment of their quality and correctness, which in turn hampers the progress of genomic research. 

This issue underscores the critical role of reproducibility in scientific research, particularly in the field of genomics. 
Formally, reproducibility refers to the ability to obtain consistent experiment results using the same input data, methods, and tools. Over the past decades, thanks to promotion by researchers~\cite{goodman2016does, begley2015reproducibility, national2019reproducibility} and the government~\cite{lusoli2020reproducibility}, there has been a growing awareness of the importance of reproducibility, and more researchers are willing to share the datasets along with their research outcomes. 
However, the sharing of genomic datasets poses significant challenges due to the sensitive nature of the data involved~\cite{lin2004genomic}.
For instance, if a genomic dataset is shared publicly, an attacker might 
infer with high confidence whether a particular victim has a specific trait or disease~\cite{zhang2017bayesian}. 
This kind of exposure represents a significant threat to personal privacy and could result in severe consequences, such as discrimination or safety risks. 

Hence, there is a crucial need for an approach to validate GWAS outcomes that (i) ensures the individuals' genomic privacy, safeguarding against state-of-the-art inference attacks, and (ii) enables validation of the research outcomes while maintaining the integrity and statistical accuracy of the data in the shared datasets. This will enable recipients (verifiers) to reproduce research outcomes and detect minor miscalculations made by researchers.

Several works focus on enabling validation of GWAS outcomes in a privacy-preserving manner.
Halimi et al.~\cite{halimi2021privacy} propose a framework that validates GWAS results, but this approach has inherent uncertainty (i.e., a grey area), making minor error detection less reliable. 
While some works focus on privacy-aware data sharing, they fall short in the genomic context. One GAN-based approach~\cite{yelmen2021creating}, proposed by Yelmen et al., attempts to generate synthetic genomic data for sharing. However, it performs poorly on real-world genomic datasets, producing synthetic data with low utility, making GWAS outcome validation nearly impossible.
DPSyn~\cite{li2021dpsyn} and PrivBayes~\cite{zhang2017privbayes} create datasets based on differentially private query results. However, these methods struggle with high-dimensional genomic data, making them impractical for datasets with thousands of features.
Other researchers have explored perturbing datasets before sharing~\cite{kasiviswanathan2011can, ji2021differentially, xiao2015protecting}. However, excessive noise often distorts the shared copies, resulting in insufficient GWAS properties to support reproducibility in practice.

In this paper, we propose PROVGEN (PRivacy-preserving Outcome Validation for GENomic Data), a novel framework that securely shares genomic datasets while ensuring reproducibility in GWAS outcome validation. We focus on datasets of point mutations in DNA, namely Single Nucleotide Polymorphisms (SNPs, introduced in Section~\ref{back:genome}), as they are the most popular ones in biomedical research \cite{mitchell2004discrepancies} and genome-wide association studies (GWAS) \cite{carlson2003additional}. Note that the shared dataset is not intended for the primary use of data (e.g., conducting research) by medical experts, due to the noise introduced by differentially-private mechanisms. In this work, we focus on the secondary use of genomic datasets, i.e., \textbf{reproducibility and validation of GWAS outcomes} by other researchers, where noise in the shared dataset can be tolerated.

The overall procedure of the proposed scheme can be described in two stages: data perturbation and utility restoration. 
In the \textbf{data perturbation} stage, genomic data is initially encoded into binary values and then perturbed by XORing it with binary noise. The probability distribution used for generating the noise is carefully calibrated by leveraging the column-wise correlation of the SNPs from publicly available datasets. The core part of the data perturbation process is the \textbf{efficient binary noise generation (EBNG)} (see Definition \ref{def:relaxed-xor}), which is an improvement of the XOR mechanism initially proposed in \cite{ji2021differentially}. The noise sampling of the original XOR mechanism is extremely time-consuming, making it impractical for large datasets such as genomic datasets. 
In contrast, our method addresses this limitation by analyzing the upper bounds of the marginal probabilities of each noise element. Notably, this efficient approach is not exclusive to binarized genomic data; it can be adapted for other relational datasets with intercorrelated entries, especially where sampling noise directly from joint distributions is computationally challenging.
In the \textbf{utility restoration} stage, we introduce a post-processing scheme aiming to improve the GWAS utility distorted by the noise introduced in the first stage. We adjust the Minor Allele Frequencies (MAFs) in our dataset to align with the MAF values provided along with the research findings by modifying certain values in our dataset. This postprocessing step significantly improves the GWAS outcomes derived from the noisy dataset without violating the differential privacy guarantee, thereby enabling reliable validation of GWAS results using the final shared dataset.

We evaluate our proposed scheme on three genomic datasets from the OpenSNP project~\cite{opensnp} with regard to its performance in GWAS outcome validation, data utility, resistance against membership inference attacks (MIAs), and time efficiency. For comparison, we implemented several alternative approaches: local differential privacy (LDP)~\cite{kasiviswanathan2011can}, GAN-AG~\cite{yelmen2021creating}, and two synthetic methods under differential privacy, DPSyn~\cite{li2021dpsyn} and PrivBayes~\cite{zhang2017privbayes}.  Our results demonstrate that PROVGEN outperforms all four methods in identifying subtle errors in GWAS outcomes. In contrast, alternative approaches either fail to detect significant errors (as in LDP~\cite{kasiviswanathan2011can}), suffer from severe utility degradation (as observed with GAN-AG~\cite{yelmen2021creating}), or face prohibitive time complexity issues (as with DPSyn~\cite{li2021dpsyn} and PrivBayes~\cite{zhang2017privbayes}). Moreover, in terms of data utility (e.g., point-wise and statistical errors) and robustness against MIAs, our scheme consistently outperforms the existing methods. Overall, the experimental results underscore the effectiveness of PROVGEN for GWAS outcome validation in real-world scenarios.

\smallskip
Our main contributions are summarized as follows.

\begin{itemize}
 \item We propose the first approach, PROVGEN, to enable GWAS reproducibility checks via privacy-preserving genomic dataset sharing.

 \item We design an innovative two-stage scheme under differential privacy that effectively detects unintentional errors in GWAS outcomes.

 \item We develop a novel XOR-based method, inspired by Ji et al.~\cite{ji2021differentially}, significantly reducing the time complexity of noise generation.

 \item We evaluate our scheme on three real-world genomic datasets and demonstrate that it outperforms existing methods in terms of accuracy, utility, and privacy.
\end{itemize}



\noindent\textbf{Roadmap.} We introduce related work in Section~\ref{sec:work}. We provide the preliminaries in Section~\ref{sec:background} and outline the system settings in Section~\ref{sec:setting}. We demonstrate the workflow in Section~\ref{sec:workflow} and elaborate on the details of our approach in Section~\ref{sec:methodology}. Evaluation results are presented in Section~\ref{sec:evaluation}. Finally, we conclude the paper in Section~\ref{sec:conclusion} and discuss its limitations and future work.

\section{Related Work}
\label{sec:work}

\subsection{Reproducibility}
\label{sec:work_repro}
Reproducibility ensures that experiment results are consistent in the same setting (e.g., using identical input data and methods), which facilitates research quality assessment~\cite{goodman2016does} and advances scientific knowledge~\cite{Lin2020Learning}. 
It has been promoted for several years by both researchers and government~\cite{begley2015reproducibility, national2019reproducibility}.
Researchers typically share datasets used in their research such that everyone can reconstruct the same experiments and validate their results. Some examples of these datasets include ImageNet~\cite{ILSVRC15} and the Iris dataset~\cite{fisher1936use}. 

However, certain datasets such as genomic and location datasets, may contain sensitive data, and thus basic anonymization techniques, e.g., hiding identifiable information, may not sufficiently protect against adversarial attacks~\cite{de2013unique}.
Halimi et al.~\cite{halimi2021privacy} propose a framework for validating GWAS results by comparing published MAFs with a noisy researcher-side MAF using a threshold derived from public genomic data. However, this method has inherent ambiguity, making the detection of small errors unreliable.
Similarly, cryptographic approaches (e.g.,~\cite{sim2020achieving}) have been explored for secure GWAS validation, but their high computational cost limits their feasibility for large-scale genomic studies.
In this paper, we address this challenge by proposing a differentially-private scheme for sharing genomic datasets while preserving high data utility. Our approach enables researchers to reproduce GWAS studies conducted on the original dataset while ensuring privacy protection for individuals.

\subsection{Privacy-Preserving Dataset Sharing}
\label{sec:work_privacy}
Differential privacy has become the gold standard for releasing aggregated statistics in a privacy-preserving way. Solutions to achieve differential privacy introduce calibrated noise to the query results, and thus prevent an attacker from gaining excessive information by observing them. Quite a few works have applied differential privacy to share/release datasets. For example, Chanyaswad et al. \cite{chanyaswad2018mvg}  apply Gaussian noises in matrix format to sanitize a numerical dataset represented as a table. Ji et al. \cite{ji2021differentially} develop the XOR mechanism to release a binary dataset by using noise attributed to the matrix-valued Bernoulli distribution. Andr\'es et al.~\cite{andres2013geo} consider the release of geolocation datasets by using Laplace noises and achieving geo-indistinguishability (a variant of differential privacy).

Meanwhile, there are also attempts to share datasets by synthesizing them under differential privacy guarantees. For example, Li et al.~\cite{li2021dpsyn} propose a scheme that generates synthetic datasets by concerning pairwise marginal distribution of features and auxiliary information. Zhang et al.~\cite{zhang2017privbayes} synthesize datasets using Bayesian networks, where conditional probabilities are noisy and protected under differential privacy. Gursoy et al.~\cite{gursoy2018differentially} release synthetic datasets of location trajectories by designing an algorithm that combines four noisy features extracted from the original dataset. Yelmen et al.~\cite{yelmen2021creating} propose a GAN-based approach, which generates artificial genomic sequences by capturing internal correlations within the original dataset. However, its utility significantly degrades as the DNA sequence length increases.
In this paper, we propose a privacy-preserving dataset sharing scheme for GWAS outcome validation that effectively detects subtle errors while maintaining high data utility. We compare our approach with existing methods and demonstrate its superior performance in Section~\ref{sec:evaluation}.

\subsection{Genomic Privacy}
\label{sec:work_genome}

Encryption-based approaches~\cite{atallah2003secure, blanton2012secure} are often inefficient concerning computational and communication costs, so the implementation of such approaches is hardly practical. Instead, differential privacy is heavily adopted. 
Uhler et al.~\cite{uhlerop2013privacy} propose a method to release GWAS statistics (e.g., $\chi^2$), and Yu et al.~\cite{yu2014scalable} improve this work by allowing an arbitrary number of case and control individuals while considering auxiliary information. Yilmaz et al.~\cite{yilmaz2022genomic} consider the correlations between SNPs and propose dependent local differential privacy to release individual genomic records. Yet, it only works for individual genomic sequences and cannot be extended to dataset sharing. Our approach publishes entire genomic datasets under differential privacy, while preserving high data utility and essential GWAS statistical properties.

\section{Preliminaries}
\label{sec:background}
In this section, we cover some basic knowledge relevant to the paper, including genomics, differential privacy and its applications. 

\subsection{Genomic Data}
\label{back:genome}
DNA (Deoxyribonucleic Acid) consists of two complementary strands forming a double-helix structure, with each strand composed of nucleotides. There are four nucleotide bases: Adenine (\textbf{A}), Thymine (\textbf{T}), Cytosine (\textbf{C}), and Guanine (\textbf{G}). They pair specifically (\textbf{A} with \textbf{T}, and \textbf{C} with \textbf{G}) to encode genetic information.

A Single Nucleotide Polymorphism (SNP) is the most common type of genetic variation, where a single nucleotide differs at a specific position in the genome among individuals. For a variation to be classified as an SNP, it must occur in at least 1\% of the population~\cite{snp}.

Each SNP can have different alleles, representing the possible nucleotide variations at a given position. The major allele is the more frequently occurring variant in a population, while the minor allele is the less common one. For example, if a specific SNP location contains \textbf{C} in most individuals but \textbf{T} in minority, then \textbf{C} is the major allele, and \textbf{T} is the minor allele. The SNP value reflects the count of minor alleles in an individual’s genome:
\begin{itemize}
    \item \textbf{0}: homozygous for the major allele
    \item \textbf{1}: heterozygous, carrying one copy of the minor allele
    \item \textbf{2}: homozygous for the minor allele
\end{itemize}

Over 600 million SNPs have been identified across global populations, and they play a crucial role in Genome-Wide Association Studies (GWAS), where researchers investigate genetic variations linked to specific diseases or traits~\cite{mittag2015influence}.

\subsection{Minor Allele Frequency (MAF)}
\label{back:maf}
The Minor Allele Frequency (MAF) represents the proportion of the less common allele (minor allele) at a specific SNP position within a population. Each individual carries two alleles at a given SNP location (one from each parent) and MAF helps quantify how frequently the minor allele appears in the population. MAF is computed as:  
\begin{equation}
    \text{MAF} = \frac{\text{Number of Minor Alleles in Population}}{\text{Total Number of Alleles in Population}}.
\end{equation}


MAF is a crucial metric in genetic research, particularly in Genome-Wide Association Studies (GWAS), as it helps determine whether genetic variations are common or rare in a population. SNPs with high MAF are considered common variants, while those with low MAF (typically below 5\%) are classified as rare variants.

\subsection{Genome-Wide Association Studies}
\label{sec:back_gwas}

Genome-wide association studies (GWAS) are a widely used approach for identifying correlations between genetic variations and specific traits or phenotypes~\cite{duncan2019genome, tam2019benefits, korte2013advantages, cantor2010prioritizing}. In a typical GWAS, individuals are divided into \textbf{case} and \textbf{control} groups based on the presence or absence of a particular characteristic, where the case group exhibits the trait, and the control group does not.  

A \textbf{contingency table} is constructed to summarize the statistical distribution of SNP values between these groups, as shown in Table~\ref{tab:cont_tab}. In this table, $S_i$ represents the number of individuals in the case group with an SNP value of $i$ at a specific genomic position, while $R_i$ denotes the corresponding count in the control group. The total number of individuals across both groups is denoted by $N$.  

\begin{table}[h]
\small
\centering
\caption{A contingency table}
\begin{tabular}{|lcccc|}
\hline
        & \multicolumn{3}{c}{Genotype} &       \\ \cline{2-4}
        & 0        & 1       & 2       & Total \\ \hline
Case    & $S_0$    & $S_1$   & $S_2$   & $S$   \\ 
Control & $R_0$    & $R_1$   & $R_2$   & $R$   \\ 
Total   & $N_0$    & $N_1$   & $N_2$   & $N$   \\
\hline
\end{tabular}
\label{tab:cont_tab}
\end{table}

For example, in a study on \textbf{lactose intolerance}, if $S_2 = 10$, it means that 10 individuals with lactose intolerance are homozygous for the minor allele at that SNP position.  
Common statistical measures used in GWAS analysis, such as the \textbf{chi-square ($\chi^2$) test} and the \textbf{odds ratio test}, are discussed in Section~\ref{sec:eval_metrics}.

\smallskip
\noindent\textbf{The typical GWAS process involves the following steps:}  
\begin{enumerate}  
    \item \textbf{Data collection:} Researchers collect genomic data from sources such as hospitals, including individuals with the disease, phenotype, or trait of interest, as well as those without the condition or characteristic.  

    \item \textbf{Group formation:} The collected data are separated into a case group (individuals with the characteristic) and a control group (individuals without the characteristic).  

    \item \textbf{Preprocessing:} The dataset is cleaned and prepared for analysis by removing unnecessary data and generating contingency tables to summarize SNP-level information.  

    \item \textbf{Statistical analysis:} Statistical tests, such as the $\chi^2$ test, are performed on the contingency tables to identify statistically significant SNPs potentially associated with the disease or trait.  

    \item \textbf{Publication of findings:} The significant SNPs and their associated statistics are published as GWAS research findings, often shared publicly to facilitate further study.  
\end{enumerate}

\subsection{Differential Privacy}
\label{sec:dp}
Differential privacy (DP) quantifies privacy and limits the inference of any single individual from observing the query results between neighboring databases. The formal definition is as follows:

\begin{mydef}[\textbf{Differential Privacy}]~\cite{dwork2008differential}
\label{def:dp}
 For any neighboring datasets $D, D'$ that differ only in one data record, a randomized algorithm $\mathcal{M}$ satisfies $\epsilon$-differential privacy if for all possible outputs $\mathcal{S} \subseteq Range(\mathcal{M})$
$$ Pr(\mathcal{M}(D) \in \mathcal{S}) \leq e^\epsilon * Pr(\mathcal{M}(D') \in \mathcal{S})\text{.}$$
\end{mydef}

The parameter $\epsilon$ quantifies the degree of privacy leakage by the algorithm $\mathcal{M}$. A smaller $\epsilon$ value signifies less privacy loss. and thus a higher level of protection, while $\epsilon = \infty$ indicates no privacy protection at all.

An important proposition of differential privacy is its \textbf{immunity to post-processing}. It ensures that the differential privacy guarantee still holds when a mapping function is performed on the output from a differentially private mechanism as long as the function does not utilize the actual value. The formal definition is as follows: 
\begin{proposition}[\textbf{Post-processing}]~\cite{dwork2014algorithmic}
\label{prop:immunity}
Let $\mathcal{M}$ be a randomized algorithm that is $\epsilon$-differentially private. For any arbitrary randomized mapping $f:\mathcal{R}^q \rightarrow \mathcal{R}^r$ where $p,q \in \mathbb{N}^+$, $f\circ \mathcal{M}$ is $\epsilon$-differentially private.

\end{proposition}

Hence, perturbations to the differentially private outputs without knowing the original values do not violate the privacy guarantee.

\subsection{The XOR Mechanism}\label{sec:xor-theorem} 
Since we utilize an improved version of it, we also revisit the definition and privacy guarantee of the XOR mechanism proposed in \cite{ji2021differentially}.
\begin{mydef}[XOR Mechanism]\label{def:eom}
Given a binary- and matrix-valued query $\eta_x$ mapping a dataset to a binary matrix, i.e.,   $\eta_x:  D \rightarrow  \{0,1\}^{n\times p}$,   the XOR mechanism is defined as
\begin{equation*}
\mathbb{XOR}(\eta_x(D),\B) = \eta_x(D)\oplus\B,
\end{equation*}
where $\oplus$ represents the XOR operator, and $\B$, a binary matrix noise within $\in\{0,1\}^{n\times p}$, follows a matrix-valued Bernoulli distribution with a quadratic exponential dependence structure, i.e.,  $$\B\sim{\rm{Ber}}_{n,p}(\T,\La_{1,2},\cdots,\La_{n-1,n}).$$ 
\end{mydef}

\subsubsection{PDF of Matrix-valued Bernoulli Distribution}\label{sec:bernoulli_pdf}
The PDF of this matrix-valued Bernoulli distribution with quadratic exponential dependency, i.e., $\B\sim{\rm{Ber}}_{n,p}(\T,\La_{1,2},\cdots,\La_{n-1,n})$  is parameterized by matrices  $\T,\La_{1,2},\cdots,\La_{n-1,n}\in\mathcal{R}^{p\times p}$ and is expressed as 

\begin{gather}\label{eq:bernoulli_pdf}
    \begin{aligned}
f_{\B}(\BB) &= C(\T,\La_{1,2},\cdots,\La_{n-1,n}) \\
\times & \exp\big\{\Tr[\BB\T\BB^T]+\sum_{i=1}^n\sum_{j\neq i}^n\Tr[\J_{ij}\BB\La_{i,j}\BB^T]\big\},
\end{aligned}
\end{gather}
where 
\begin{gather*}
\begin{aligned}
&C(\T,\La_{1,2},\cdots,\La_{n-1,n})=\\
&\resizebox{0.98\hsize}{!}{$\Big[\sum_{\BB_k}\exp\Big\{\Tr[\BB_k \T \BB_k^T]+\textstyle\sum_{i=1}^n\sum_{j\neq i}^n\Tr[\J_{ij} \BB_k\La_{i,j} \BB_k^T]\Big\}\Big]^{-1},$}
\end{aligned}
\label{norm_const}
\end{gather*}
is the normalization constant,   
  $\BB_k\in\{0,1\}^{n\times p}$, and $\J_{ij}$ is the  matrix of order $n\times n$ with $1$ at the $(i,j)$-th position and 0 elsewhere.

    
Similar to the classical differential privacy output perturbation mechanisms (such as Gaussian or Laplace mechanisms), which attain privacy guarantees by constraining the parameters of the considered distributions (i.e., Gaussian or Laplace distribution), the XOR mechanism also ensures privacy by controlling the parameters in the distribution in (\ref{eq:bernoulli_pdf}). The sufficient condition for the XOR mechanism to achieve $\epsilon$-differential privacy is recalled as follows.

\begin{theorem}\label{thm:xor-privacy}
The XOR mechanism achieves   $\epsilon$-differential privacy of a matrix-valued binary query  if $\T$ and $\La_{i,j}$ satisfy 

\begin{equation}
s_f\Big(||\boldsymbol{\lambda}(\T)||_2    +   \textstyle\sum_{i=1}^{N-1}\sum_{j= i+1}^N  ||\boldsymbol{\lambda}(\La_{i,j})||_2  \Big)\leq \epsilon, 
\label{condition_general}
\end{equation}
where $s_f$ is the sensitivity of the binary- and matrix-valued query, and $||\boldsymbol{\lambda}(\T)||_2$ and $||\boldsymbol{\lambda}(\La_{i,j})||_2$ are the $l_2$ norm of the vectors composed of eigenvalues of $\T$ and $\La_{i,j}$, respectively. 
\label{eo_theorem}
\end{theorem}
\noindent In Section \ref{sec:perturbation}, we will discuss how to obtain $s_f$ when considering a binarized genomic dataset.

In practice, it is computationally prohibitive to evaluate the normalization constant (see (\ref{eq:bernoulli_pdf})) in the PDF of the matrix-valued Bernoulli distribution; thus, to generate a sample from it, \cite{ji2021differentially} resorts to the Exact Hamiltonian Monte Carlo scheme. However, this scheme is impractical for our study due to its extreme high time complexity caused by thousands of SNPs in the dataset. To address this issue, we propose a new noise generation scheme without compromising the privacy of the original XOR mechanism. In particular, each element in the noise matrix is generated using its calibrated marginal distribution. More details are deferred to Section \ref{sec:noise-generate}.


\section{System Settings}
\label{sec:setting}
In this section, we introduce the system model and the threat model.

\subsection{System Model}
\label{sec:system}
In our system, we consider a scenario in which researchers are conducting genome-wide association studies (GWAS). In this setting, we consider two key parties: \textbf{the researcher} and \textbf{the verifier}. The researcher performs GWAS on a local genomic dataset and publishes the research findings publicly. 
Typically, GWAS research findings include the name and methodology of the experiment, the number of samples in the case group, details about the control group (which may include aggregated statistics of the control group), the SNP identifier (commonly referred to as the \textbf{rsID}), and the minor allele frequency (MAF) of SNPs that are significantly associated with the trait or disease of interest.

To enhance the credibility of its research and reputation in the research community, the researcher would share the entire research dataset along with the findings for external validation. However, direct sharing of genomic datasets raises privacy concerns~\cite{lin2004genomic}. Instead, the researcher sanitizes the dataset using a privacy-preserving scheme before sharing it.

Throughout the research process, the researcher may unintentionally introduce errors, including errors during: (i) data cleaning and preprocessing, (ii) statistical analysis, or (iii) publication of findings. Our work focuses on detecting unintentional errors that occur during these stages, but excludes errors introduced during data collection. Errors at this stage are generally considered hard to detect and have not been effectively addressed by existing methods. This is widely acknowledged in the field, and no current work has provided a robust solution for identifying or correcting such errors.

Importantly, multiple errors may arise at the same or different stages. Since such errors are generally independent, we treat them as independent events. Our approach is capable of detecting such compounded errors because they collectively lead to discrepancies in GWAS results, making it easier for our scheme to identify them.

The verifier, assumed to be a peer reviewer or another researcher, seeks to validate the published findings. Using the shared dataset, the verification process proceeds as follows. First, the verifier reproduces the same GWAS experiments reported by the researcher using the shared dataset and obtains the $p$-values for all SNPs claimed as significant in the research findings. Then, the verifier calculates the percentage of SNPs reported as significant in the original findings that remain statistically significant in the reproduced results (possibly using a relaxed $p$-value threshold instead of the original $p$-value), which is termed the \textbf{SNP retention rate}.

To assess the trustworthiness of the findings, the verifier compares the SNP retention rate from the reproduced results to a theoretical ideal or expected rate, which represents the anticipated retention rate under error-free conditions. The primary goal of this paper is to ensure that meaningful differences in retention rates can be observed, enabling the verifier to detect potential errors. Note that the estimation of such an expected rate, which can be obtained by using additional auxiliary datasets, is beyond the scope of this paper. If this difference falls within a specified threshold, the findings are deemed reliable. Otherwise, the verifier may initiate further investigation or request additional detailed information, pending Institutional Review Board (IRB) approval.

\subsection{Threat Model}
\label{sec:threat}

In our framework, we assume that the researcher is honest yet cautious, holding the original genomic dataset without sharing it directly. Meanwhile, an honest researcher may still unintentionally provide incorrect GWAS outcomes due to computational errors, which could mislead other researchers. Our scheme aims to address this issue by offering a means to reproduce and validate GWAS experiments, enhancing their reliability.

Note that our scope does not extend to scenarios involving a malicious researcher who might intentionally fabricate datasets to report false results. Deliberately creating and using synthetic datasets to produce inaccurate findings poses a challenge that is nearly impossible in most data analysis contexts, not only in GWAS. In this case, only those with direct access to the original data can validate the authenticity of research finding. Moreover, the ethical implications and potential consequence of using fabricated dataset (e.g., damage to the researcher's credibility and reputation from funding agencies) serve as strong deterrents against such misconduct. 

The verifier may be malicious and curious about the original dataset. Such a malicious verifier, acting as an attacker, may conduct membership inference attacks (MIAs) to determine if an individual (the victim) is part of the shared dataset or not. Since individuals in a genomic dataset often share an attribute (e.g., trait or disease), linking the victim to the dataset could also associate them with that attribute.
For instance, assume the researcher shares a dataset consisting of heart disease patients. An attacker could use the published research findings and the shared dataset to predict the target's presence in the dataset. If the analysis indicates that the target is likely a member, the attacker could infer a potential association of the individual with heart-related diseases.

We assume that the attacker has access to two key pieces of information: (i) the shared dataset from the researcher and (ii) the specific trait/disease of the individuals in the dataset, e.g., heart disease in the previous example. In addition, the attacker can exploit auxiliary knowledge to launch MIAs by constructing a reference dataset of individuals without the trait/disease (e.g., from the 1000 Genomes project~\cite{1000genome}). 
We consider the following MIAs: 1) Hamming distance-based test (HDT)~\cite{halimi2021privacy}, 2) decision tree, 3) random forest, 4) XGBoost~\cite{chen2016xgboost}, 6) Support Vector Machine~\cite{cortes1995support}, and 7) neural network. More details will be  deferred to Section~\ref{eval:mia_metric}. 

\section{System Workflow of PROVGEN}\label{sec:workflow}

\begin{figure*}[ht]
 \centering

  \includegraphics[width=\textwidth]{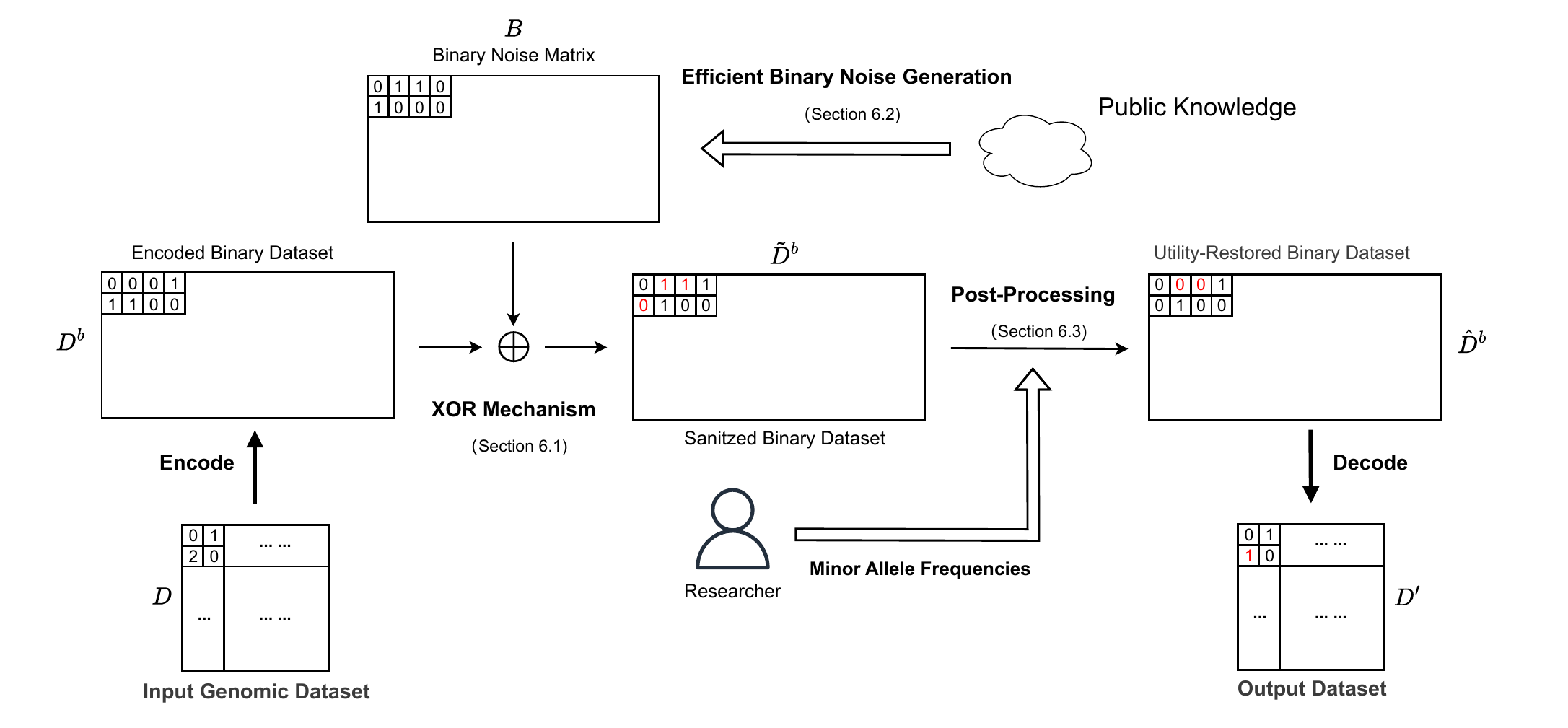}

\caption{
The workflow of PROVGEN operates as follows: 1) The input dataset $D$ is encoded into a binary form $D^b$ and subjected to an XOR operation with binary noise, generated through Efficient Binary Noise Generation (EBNG). 2) We utilize the Minor Allele Frequencies (MAFs) of SNPs that are published in the research findings to enhance the data utility of the noisy dataset $\hat{D}^b$ using optimal transport. Finally, we convert the optimized binary dataset $\hat{D}^b$ back into its original SNP format to obtain the final shared dataset $D'$.
}

\label{fig:workflow}
\end{figure*}

Our proposed workflow is depicted in Figure~\ref{fig:workflow}.

During the data perturbation stage, we first encode the genomic dataset into a binary matrix. Each SNP is represented using two bits while considering the biological property (discussed in Section \ref{sec:perturbation}). We then implement a noise sampling scheme, an adaptation of the XOR mechanism~\cite{ji2021differentially}, optimized for efficient generation of large datasets. This scheme generates a noisy version of the binary matrix while considering inherent correlations among SNPs from publicly available datasets.  

In the utility restoration stage, we address the utility degradation caused by noise addition. We develop a post-processing technique focused on enhancing the GWAS utility distorted in the first stage. This involves aligning the Minor Allele Frequencies (MAFs) in the noisy dataset with those that have been published in research findings by flipping allele values using optimal transport~\cite{intro_to_ot}. Following this, we convert the altered dataset back into genomic space and make it available to verifiers for validation.

\section{Methodology}
\label{sec:methodology}

In this section, we introduce the technical details of our proposed scheme.
We discuss how to generate a correlation-aware noise matrix in Section~\ref{sec:perturbation} and present our noise generation scheme in Section \ref{sec:noise-generate}.
In Section~\ref{sec:method_postprocessing}, we explain how we restore GWAS utility using MAF.

\subsection{Genomic Dataset Perturbation}
\label{sec:perturbation}

Existing methods are not effective for genomic data due to two reasons: they either exhibit high time complexity~\cite{li2021dpsyn,zhang2017privbayes} or fail to appropriately address the inherited correlation among SNPs~\cite{kasiviswanathan2011can}, leading to significant utility loss, as evidenced by our preliminary experiments. We overcome these challenges by converting genomic datasets into binary space. Specifically, we encode each SNP value to 2 bits according to the conversion metric shown in Table~\ref{tab:encoding} and generate a binary version of the genomic dataset $D^b \in \{0,1\}^{n\times 2m}$. It is important to note that this binary representation of SNPs is consistent with their biological characteristics (refer to Section~\ref{back:genome}). As detailed in Section~\ref{sec:back_gwas}, SNP data have three values (0, 1 and 2) that indicate the number of minor alleles in a gene. Each allele, inherited from one parent, contributes to the SNP value: `00' for value 0 (no minor allele), `01' for value 1 (one minor allele), and `11' for value 2 (both parents with a minor allele). The binary matrix resulting from this encoding effectively simulates the allele distribution in the genomic sequence. Therefore, flipping one binary value in the binary dataset is analogous to flipping one allele, thus maintaining biological consistency in our data representation.

\begin{table}[h]
\centering
\caption{Conversion between SNP values and binary format.}
\label{tab:encoding}
\begin{tabular}{cc}
\toprule
\textbf{Genomic Value} & \textbf{Binary Format} \\
\midrule
0 & \texttt{00} \\
1 & \texttt{01} \\
2 & \texttt{11} \\
\bottomrule
\end{tabular}
\end{table}

After encoding, we implement the XOR mechanism~\cite{ji2021differentially} to perturb the binary-encoded SNP dataset. The perturbation is represented as $\tilde{D}^b = D^b\oplus B$, where $D^b$ is the original binary SNP dataset, and $\tilde{D}^b$ is the perturbed outcome. The operator $\oplus$ denotes the (exclusive or) XOR operation, and $B\in\{0,1\}^{n\times 2m}$ is the binary noise matrix, sampled from the matrix-valued Bernoulli distribution.  

The parameters of this distribution are calibrated with respect to the privacy parameter $\epsilon$ and the sensitivity $s_f$ of the binary encoding between $D$ and $\hat{D}$, where $D\sim \hat{D}$ denotes neighboring genomic datasets that differ by a single individual's genomic record. Mathematically, the sensitivity is defined as:  
\begin{equation}
  s_f = \sup_{D,\hat{D}} \Vert D^b \oplus {\hat{D}}^b \Vert_F^2.
\end{equation}
Here, $\Vert \cdot \Vert_F^2$ represents the Frobenius norm, and $s_f$ quantifies the maximum number of differing entries between $D^b$ and $\hat{D}^b$.  

\smallskip
\textbf{Motivation of Adopting XOR Mechanism.}
Two reasons justify the suitability of the XOR mechanism for our scenario: (i) it can directly generate a binary noise matrix attributed to the matrix-valued Bernoulli distribution with a quadratic exponential dependence structure~\cite{lovison2006matrix}; and (ii) the dependence structure enables the characterization of various correlations among the encoded SNPs.  

It is noteworthy that, in practice, there are infinite ways to generate the values of $\T$ and $\La_{ij}$ (i.e., the parametric matrices in the XOR mechanism). As long as the sufficient condition in (\ref{condition_general}) is satisfied, achieving $\epsilon$-differential privacy is possible. In \cite{ji2021differentially}, the authors assume that $\T$ and $\La_{ij}$ are positive definite matrices and suggest generating them by solving a complex optimization problem. In this work, we relax this assumption and generate $\T$ by considering the biological characteristics of SNPs.  

In particular, the entries of $\T$ (cf. (\ref{eq:bernoulli_pdf}) in Section~\ref{sec:bernoulli_pdf}) are referred to as ``feature-association'' values~\cite{ji2021differentially,lovison2006matrix}. They model the correlations between the columns of $D^b\in\{0,1\}^{n\times 2m}$, representing the inherent dependencies among SNPs~\cite{naveed2015privacy}. To preserve these inherent correlations, we utilize a publicly available reference dataset that contains the same set of SNPs as $D$ (e.g., the control group in a case-control study). Such datasets can be obtained from open-source genomic repositories, such as 1000 Genomes~\cite{1000genome}.  

To model these correlations, we employ a \textbf{log-linear association approach}, constructing a matrix $\widetilde{\T}\in\mathbb{R}^{2m\times 2m}$ with diagonal entries $\widetilde{\theta}_{p,p}$ and non-diagonal entries $\widetilde{\theta}_{p,q}$ ($p\neq q$), computed as:  
\begin{gather}
\begin{aligned}
& \widetilde{\theta}_{p,p} = \log \frac{\Pr(M_{p}=0)}{\Pr(M_{p}=1)}, \quad p\in[1,2m],\\
& \widetilde{\theta}_{p,q}= \log \frac{\Pr(M_{p}=0,M_{q}=1)\Pr(M_{p}=1,M_{q}=0)}{\Pr(M_{p}=1,M_{q}=1)\Pr(M_{p}=0,M_{q}=0)}, 
\end{aligned}
\label{eq:empirical_theta}
\end{gather}
where $p,q\in[1,2m], p\neq q$,  
$M$ is the binarized version of a publicly available dataset, $\Pr(M_{p}=0)$ represents the frequency of values equal to $0$ in the $p$-th column, and $\Pr(M_{p}=0,M_{q}=1)$ denotes the frequency of tuples where the value is $0$ in the $p$-th column and $1$ in the $q$-th column.  

The entries of $\La_{ij}$, referred to as ``object-association'' values, model the correlations between rows $i$ and $j$ of $D^b$, representing kinship relationships~\cite{ji-btac243}. If the genomic dataset contains SNP sequences of family members, $\La_{ij}$ can be derived using Mendel's law. Given the privacy parameter $\epsilon$, we determine the distribution parameters ($\T$ and $\La_{ij}$) based on correlations obtained from publicly available genomic datasets of the same nature. In this scenario, $\La_{ij}$ is generated using the log-linear association approach to preserve kinship-based correlations.

\smallskip
By invoking Theorem~\ref{thm:xor-privacy}, we have 
\begin{equation}
s_f \big(||\boldsymbol{\lambda}(\T)||_2 +   \textstyle\sum_{i=1}^{n-1}\sum_{j= i+1}^n  ||\boldsymbol{\lambda}(\La_{i,j})||_2 \big)   \leq \epsilon_x, 
\label{condition_general_T}
\end{equation}
as a sufficient condition to protect the privacy of genomic dataset without kinship correlations.

\subsection{Efficient Binary Noise Generation}
\label{sec:noise-generate}
Since the original noise generation of the XOR mechanism is time consuming and impractical for genomic datasets, we introduce a novel scheme, named \textbf{efficient binary noise generation} (EBNG). This scheme is designed to speed up the noise generation process, making it feasible   for use with large-scale genomic data.

First, we review the following lemma which connects the matrix-valued Bernoulli distribution with the multivariate Bernoulli distribution.
\begin{lemma}\cite{lovison2006matrix}\label{lemma:ber} If $\B\sim {\rm{Ber}}_{n,p}(\T,\La_{1,2},\cdots,\La_{n-1,n})$, then $\bb={\rm{vec}}(\B^T)\in\{0,1\}^{np\times 1}$ is attributed to  a  multivariate Bernoulli distribution with parameter $\bm{\Pi}$, i.e., ${\rm{vec}}(\B^T)\sim {\rm{Ber}}_{np}(\bm{\Pi})$, and  
\begin{equation*}\label{eq:pdf-bernoulli-vector}
f_{{\rm{vec}}(\B^T)} ({\rm{vec}}(\B^T)=\bb) = C(\bm{\Pi})\exp\{\bb^T\bm{\Pi}\bb\}.
\end{equation*}
The parameter $\bm{\Pi}$ and the normalization constant $C(\bm{\Pi})$ are 
\begin{equation}\label{eq:normalization-constant-vector}
\begin{aligned}
   & \bm{\Pi} = \mathbf{I}_n\otimes\T+\sum_{i=1}^n\sum_{j\neq i}^n\J_{ij}\otimes\La_{i,j},\\
 \text{and}\quad&   C(\bm{\Pi}) = \Big[   \sum_{\bb\in\mathcal{S}}   \exp\{   \bb^T \bm{\Pi} \bb   \}    \Big]^{-1},\quad \mathcal{S}=\{0,1\}^{np\times 1},
\end{aligned}
\end{equation} 
where $\otimes$ is the Kronecker tensor product.
\end{lemma}

We can adopt a simple form of the multivariate Bernoulli distribution, i.e., 
\begin{equation}\label{eq:simple-pi}
 f_b(b=\bb) = C(\bm{\Pi}) \exp\{\bb^T\bm{\Pi}\bb\}, \  \bm{\Pi} = \mathbf{I}_n\otimes\T.
\end{equation}

Instead of generating matrix-valued binary noise from $\B\sim {\rm{Ber}}_{n,p}(\T,\La_{1,2},\cdots,\La_{n-1,n})$, we generate a vectorized version from ${\rm{vec}}(\B^T)\sim {\rm{Ber}}_{np}(\bm{\Pi})$ (with the probability density function shown in (\ref{eq:simple-pi})), and then reshape the vector noise back into matrix format to perturb the encoded genomic dataset. However, the normalization constant $C(\bm{\Pi})$ (which is a summation of $2^{nm}$ values) is still intractable. Thus, to avoid the time-consuming Hamiltonian Monte Carlo-based sampling scheme (adopted in \cite{ji2021differentially}), we generate each element of the noise vector separately, based on its approximate marginal PDF. In what follows, we first present the noise generation scheme along with the privacy guarantee. The detailed proof is deferred to Appendix~\ref{app:proof-relaxed-xor}.

\begin{mydef}[Efficient Binary Noise Generation]
\label{def:relaxed-xor}
Given a genomic dataset $D$, suppose that each individual has $m$ SNPs (i.e., $P = 2m$ SNP bits after encoding). The efficient binary noise generation scheme perturbs the $u$-th bit ($u\in[1,2m]$) for each individual with a random binary bit $B_u$, and $\Pr[B_u=1]$ is calibrated based on the correlation among the SNP bits.  
\end{mydef}

A sufficient condition for the above perturbation to preserve $\epsilon$-differential privacy on the entire genomic dataset $D$ is shown below. 

\begin{theorem}
\label{thm:relaxed-xor}
Let $\bm{\Pi}$ be the parameter determined in (\ref{eq:normalization-constant-vector}), 
${\rm SUM}(\bm{\Pi}_u)$ is the summation of the $u$th row of $\bm{\Pi}$, and $\bm{\Pi}_{u,u}$ is the $u$th diagonal entry of $\bm{\Pi}$. Define 
\begin{equation*}
    \kappa_u = 2\times {\rm SUM}(\bm{\Pi}_u)-\bm{\Pi}_{u,u},
\end{equation*}
then, Definition \ref{def:relaxed-xor} achieves $\epsilon$-differential privacy if 
\begin{equation*}
    \Pr[B_u = 1] = \begin{cases} 
      \frac{1}{2} & \textit{if}\quad \kappa_u > ||\lambda(\T)||_2 \\
  \frac{1}{1+\exp(\kappa_u)} & \textit{if}\quad \kappa_u \leq ||\lambda(\T)||_2
   \end{cases}.
\end{equation*}
\end{theorem}

\subsection{Restoring GWAS Utility via Post-Processing }
\label{sec:method_postprocessing}

Efficient binary noise generation introduces an excessive amount of noise into the encoded genomic dataset. This poses a significant challenge for verifiers, as the resultant dataset from the XOR mechanism compromises reliable GWAS outcome validation due to substantial utility loss. To address this issue, we employ a post-processing strategy that leverages Minor Allele Frequencies (MAFs) shared by the researcher in the research findings to enhance the dataset's utility. 
Notably, sharing of MAF values is permitted under the NIH Genomic Data Sharing (GDS) policy~\cite{nih-gds} and is commonly included in recent research findings.

In our approach, the published MAFs are represented as  
\begin{equation}
    \mathcal{M}^r=\{\mathcal{M}^r_0, \mathcal{M}^r_1, \dots, \mathcal{M}^r_m\},
\end{equation}  
where $\mathcal{M}^r_j$ denotes the MAF of the $j$-th SNP in the target dataset. We also compute the MAFs, denoted as $\tilde{\mathcal{M}}$, from the dataset generated in the first stage.

Our objective is to align the MAF distribution of the noisy dataset $\tilde{D}^b$ with the published MAFs while mitigating utility degradation, namely, minimizing the number of allele flips required. The post-processing procedure for each SNP $j$ follows these steps:
\begin{enumerate}
    \item Compute the MAF value $\tilde{\mathcal{M}}_j$ of the SNPs in the noisy (binarized) dataset $\tilde{D}^b$.
    \item Apply an optimal transport approach, specifically the earth mover's distance~\cite{rubner1998metric}, to transition from $\tilde{\mathcal{M}}_j$ to the reference MAF $\mathcal{M}^r_j$. This approach determines the percentage of alleles that need to be flipped and their corresponding values.
    \item Determine the exact number of alleles to be flipped by applying the floor function to the product of the computed percentage and the total number of alleles.
    \item Randomly select the determined number of alleles and flip them.
\end{enumerate}

Further details regarding this optimization problem can be found in Appendix~\ref{sec:method_postprocessing}.

\smallskip
After post-processing, we convert the dataset back into genomic space following Table~\ref{tab:encoding}. During this step, any invalid binary output (e.g., ``\textbf{10}'') is converted to ``\textbf{1}'' in SNP representation. This adjustment aligns with biological properties and enhances data utility while maintaining the integrity of the original genomic information. Meanwhile, this correction does not violate privacy guarantees, as differential privacy is immune to post-processing as shown in Proposition~\ref{prop:immunity}.

\medskip
\textbf{End-to-end Privacy Guarantee.} The efficient binary noise generation satisfies $\epsilon$-differential privacy based on Theorem \ref{thm:relaxed-xor}. According to Proposition~\ref{prop:immunity}, differential privacy is immune to post-processing if the post-processing step does not utilize the private dataset $D$. In our approach, GWAS utility is restored using Minor Allele Frequencies (MAFs) that are already publicly shared in the research findings, which means the privacy guarantee holds and no additional privacy risk is introduced. Therefore, our entire sharing scheme remains $\epsilon$-differentially private.

\section{Evaluation}
\label{sec:evaluation}
We conducted a comprehensive evaluation of our scheme using three real-world genomic datasets from the OpenSNP project~\cite{opensnp}. For comparison, we introduced a local differential privacy approach based on randomized response~\cite{kasiviswanathan2011can} as a baseline. Additionally, we evaluated the state-of-the-art synthesis-based approach~\cite{yelmen2021creating} for genomic dataset sharing to highlight its limitations on realistic genomic datasets.  

Furthermore, we implemented two widely used synthetic data generation approaches, DPSyn~\cite{li2021dpsyn} and PrivBayes~\cite{zhang2017privbayes}, both of which were winners of the \textit{NIST Differential Privacy Synthetic Data Challenge}~\cite{nist}. These methods are commonly used for tabular data and serve as alternative solutions when GAN based approaches fail for genomic dataset sharing. However, as discussed, these synthesis based methods suffer from significant computational complexity and can only handle genomic datasets with approximately 130 SNPs. This is far below the millions of SNPs typically present in an individual's genome.

We evaluated our scheme across four key dimensions: GWAS outcome validation (Section~\ref{sec:eval_gwas}), data utility (Section~\ref{sec:eval_utility}), resistance to membership inference attacks (MIAs) (Section~\ref{sec:eval_privacy}), and time efficiency (Section~\ref{sec:time}). The evaluation metrics used in our analysis are detailed in Section~\ref{sec:eval_metrics}.

\subsection{Datasets}
\label{sec:eval_dataset}

We leveraged the OpenSNP project~\cite{opensnp} to construct sample datasets for evaluation. We select three phenotypes, i.e., lactose intolerance, hair color, and eye color, and introduce the specifications in Table~\ref{tab:dataset}.
We constructed reference datasets for each phenotype, ensuring that the SNPs aligned with those in the target dataset.

\begin{table}[ht]
\begin{tabular}{|c|c|c|}
\hline
\textbf{Phenotype}           & \textbf{\# of SNPs} & \textbf{\# of Individuals} \\
\hline
Lactose Intolerance & 9091       & 60                \\
\hline
Hair Color          & 9686       & 60                \\
\hline
Eye Color           & 28396      & 401    \\           
\hline
\end{tabular}
\caption{Dataset specifications.}
\label{tab:dataset}
\end{table}

\subsection{Ethical Considerations}
In our study, we utilized existing public genomic datasets to extract inherent correlations between SNPs in the noise generation step (Section~\ref{sec:noise-generate}) and leveraged the published minor allele frequencies (MAFs) to perform  utility restoration (Section~\ref{sec:method_postprocessing}). While these datasets are publicly accessible and have been previously cleared for use in research, we recognize the importance of addressing ethical considerations in our work.

In particular, we ensure that the use of these datasets aligns with their intended purposes as defined by the original data providers. Our methodologies and objectives are consistent with the terms under which these datasets were made public. Our approach, focuses on maintaining the anonymity integral to these datasets, avoids any attempts at re-identification and strictly follows protocols to prevent de-anonymization. Meanwhile, our work does not involve direct interaction with human participants, thus significantly reducing ethical risks commonly associated with primary data collection. We adhere to recognized standards for secondary data usage, and continually stay informed about ethical guidelines and best practices in genomic research to ensure ongoing compliance. 


\subsection{Evaluation Metrics}
\label{sec:eval_metrics}
In this section, we introduce evaluation metrics regarding GWAS outcome validation, data utility, resistance against MIAs, and time complexity.

As introduced in Section~\ref{sec:system}, we use the SNP retention rate to assess the reliability of GWAS findings. This rate quantifies the proportion of significant SNPs from the original study that remain significant when reproduced using the shared dataset. Since we do not prescribe a specific threshold for acceptability, our evaluation focuses on the difference in SNP retention rates between error-free and error-injected scenarios. A difference greater than zero that shows clear separation across the x-axis (e.g., error rates) indicates the presence of potential errors. Larger deviations correspond to higher confidence in error detection. To explore how such deviations arise, we model two representative types of errors commonly encountered in GWAS pipelines.

\subsubsection{GWAS Outcome Validation}
\label{sec:eval_gwas_validation}

In an honest research setting, a researcher may inadvertently report inaccurate SNPs as part of GWAS outcomes. We categorize such errors into two types and model them accordingly. Unintentional errors may occur in many other ways as well, but we use the below simple scenarios to illustrate the consequences of such errors. 

\smallskip
\textbf{Flipping error} occurs when some $p$-values are calculated incorrectly. This type of error may arise from mistakes during any of the five stages mentioned in Section~\ref{sec:back_gwas}: data collection, group formation, preprocessing, statistical analysis, or publication of findings. We model flipping errors by randomly selecting a portion of the $p$-values and replacing them with values between 0 and 1. The error rate, denoted by $\delta_f$, parameterizes this error. The researcher correctly reports $1-\delta_f$ of the truly significant SNPs, while the remaining $\delta_f$ are reported incorrectly. 

\textbf{Noise error} happens during all stages except for the publication of findings. This type of error introduces noise into contingency tables or the calculated $p$-values. We model noise errors by adding normally distributed noise to each $p$-value reported in the research findings. The error rate $\delta_n$ denotes the scale (standard deviation) of the normal distribution from which the noise is sampled.

\smallskip
For GWAS, we focus on two statistics: the $\chi^2$ test and the odds ratio test. The $\chi^2$ test, a statistical method, assesses the significance of differences in frequencies between two groups. Specifically, it computes a $\chi^2$ statistic that reflects the discrepancy between observed and expected frequencies, i.e., in the shared dataset and the reference dataset, respectively, under the null hypothesis of no association between the phenotype/disease and the SNP position. A higher $\chi^2$ value indicates a greater deviation from the null hypothesis, suggesting a significant association. 

The odds ratio test, on the other hand, quantifies associations between a phenotype/disease and a specific SNP position using the ratio of the odds. We first calculate the odds ratio as shown in (~\ref{eq:or}).
\begin{equation}
    OR = \frac{(S_1+S_2)/(R_1+R_2)}{S_0/R_0} =\frac{R_0 (S_1+S_2)}{S_0  (R_1 + R_2)}.
    \label{eq:or}
\end{equation}
Next, we calculate the $95\%$ confidence interval of the odds ratio using the formula $\exp(\ln{(OR)}\pm 1.96\times SE(\ln{(OR)}))$, where $SE(\ln{(OR)}) = \sqrt{\frac{1}{S_1+S_2}+\frac{1}{S_0}+\frac{1}{R_1+R_2}+\frac{1}{R_0}}$. In this context, $\exp(\cdot)$ represents the exponential function and $SE(\cdot)$ is the standard error of the log odds ratio. Following this, the standard normal deviation, or $z$-value, is computed as $\frac{\ln{(OR)}}{SE\{\ln{(OR)}\}}$, and the $p$-value is the area of the normal distribution outside $\pm z$.

\subsubsection{Data Utility}
\label{sec:utility_metrics}
In addition to our primary goal, we evaluated the performance of the proposed scheme using data utility metrics for various data analysis tasks. We employed metrics including average point error, average sample error, mean error, and variance error. These metrics quantify the extent to which the integrity and statistical properties of the original dataset are preserved when shared with other researchers. The following sections detail each metric.

\paragraph{Average Point Error}  
Average point error measures the entry-level discrepancy between two datasets. Given two datasets, $D$ and $D^*$, with dimensions $n \times m$, we compute the number of mismatched entries and define the point error as:
\begin{equation}
    Error_p = \frac{\sum_{i=1}^{n} \sum_{j=1}^{m} \mathbf{1}_{D_{ij} \neq D^*_{ij}}}{n \times m}.
\end{equation}

\paragraph{Average Sample Error}  
Average sample error quantifies the distance between corresponding samples in two datasets. We calculate this using the $l_1$ norm, which represents the sum of absolute differences between each sample pair, normalized by the number of SNPs in the datasets:
\begin{equation}
    Error_s = \frac{\sum_{i=1}^{n} \sum_{j=1}^{m} |D_{ij} - D^*_{ij}|}{n \times m}.
\end{equation}

\paragraph{Mean and Variance Error}  
In addition to the previous metrics, we assess the mean and variance errors, which are key indicators of data fidelity. These errors are measured as the absolute differences in mean and variance values within the SNP domain.

\subsubsection{Resistance Against Membership Inference Attacks (MIAs)}
\label{eval:mia_metric}
We conducted a privacy evaluation against membership inference attacks (MIAs) following existing work on genomic data~\cite{halimi2021privacy}. We adhered to the threat model in Section~\ref{sec:threat} and employed multiple attacks, including the Hamming distance test and multiple machine learning-based MIAs. We do not include likelihood ratio tests in our evaluation, as done in \cite{halimi2021privacy}, because they do not involve dataset sharing \cite{sankararaman2009genomic}. We use accuracy as our evaluation metric. For each machine learning-based MIA, we train a separate model using a balanced sample of individuals from both the shared and a reference dataset. Accuracy is then evaluated using the remaining individuals in the original, unperturbed version of the shared dataset. The details of the attack strategies are outlined below.

\medskip

\noindent\textbf{Hamming Distance Test (HDT)~\cite{halimi2021privacy}.}
\label{eval:hdt_metric}
The Hamming distance test (HDT) is considered one of the most powerful MIAs against genomic datasets~\cite{halimi2021privacy}. It leverages the pairwise Hamming distance between genomic sequences in the case group (received from the researcher and subjected to perturbation for privacy guarantees) and in the control group (constructed from publicly available datasets). Specifically, for each individual $i$ in the shared dataset (the case group), a malicious client calculates the Hamming distances between individual $i$ and all individuals in the reference dataset (the control group), then records the minimum Hamming distance for individual $i$. The malicious client collects all these minimum Hamming distance for individuals in the case group and selects a threshold $\gamma$ following $5\%$ false positive rate. When attempting to identify a victim, the malicious client calculates the minimum Hamming distance between the victim's sequence and all individuals in the control group. If the minimum Hamming distance is lower than the threshold $\gamma$, the target is considered a member of the case group, and vice versa.

\noindent\textbf{Machine Learning-Based MIAs.}
\label{eval:ml_metric}
We assume that the attacker employs the following machine learning-based membership inference attacks (MIAs) for inference: decision tree (DT), random forest (RF)~\cite{pal2005random}, XGBoost~\cite{chen2016xgboost}, support vector machine (SVM)~\cite{noble2006support}, and neural networks. Upsampling methods, such as SMOTE~\cite{chawla2002smote}, are not suitable in this case due to the limited sample size (e.g., only 60 samples for the lactose intolerance dataset and the hair color dataset), which prevents generating sufficient synthetic samples.

Given the limited sample size (at most 401 samples), our experiment may not fully capture the robustness of the scheme against machine learning-based MIAs in realistic scenarios, particularly for the lactose intolerance and hair color datasets. Nonetheless, the results provide valuable insights into privacy protection. The most realistic evaluation is conducted on the eye color dataset, which contains 401 samples, offering a scenario closer to real-world settings.

\subsubsection{Time Complexity Evaluation}. To evaluate the computational efficiency of each method, we measure the time required to generate privacy-preserving datasets across a range of SNP counts. This metric is used to compare our approach with Local Differential Privacy (LDP)\cite{kasiviswanathan2011can}, DPSyn\cite{li2021dpsyn}, PrivBayes~\cite{zhang2017privbayes}, and GAN~\cite{yelmen2021creating}, using the lactose intolerance dataset as a benchmark. For each configuration, we repeat the experiment 10 times and report the average runtime to reduce the effect of variability. By varying the number of SNPs, we assess how the time consumption of each method scales with data dimensionality, thereby revealing their practicality for large-scale genomic data sharing.

\subsection{Experiment Setup}
\label{sec:eval_para}
In our experiments, we evaluate real-world genomic datasets that do not contain family members (Section~\ref{sec:eval_dataset}). Given this characteristic, we set the parametric matrices $\La_{ij} = 0$, as kinship correlations are not relevant in this scenario~\cite{lovison2006matrix}. This setting ensures that our perturbation mechanism focuses on preserving privacy while maintaining utility without unnecessary complexity. 

In this case, we only need to calibrate the value of $\T$ to satisfy (\ref{condition_general_T}). By setting  
\begin{equation}\label{eq:theta-calib}
    \T = \frac{\epsilon_x}{s_f||\widetilde{\T}||_F}\widetilde{\T},
\end{equation}
one can verify that the sufficient condition in (\ref{condition_general_T}) is satisfied. The values of $\widetilde{\T}$ are determined using a publicly available dataset via (\ref{eq:empirical_theta}).

It is important to note that we are sharing a genomic dataset containing over 9,000 SNPs, resulting in a sensitivity of 18,000 after encoding. Since our scheme operates on the entire dataset while LDP~\cite{kasiviswanathan2011can} applies to individual SNPs, a direct comparison using the same privacy parameter is not feasible. To ensure a fair evaluation, we follow similar works~\cite{halimi2021privacy} and introduce the concept of ``effective $\epsilon$,'' denoted as $\epsilon_e$, which represents the actual privacy budget allocated per SNP. By adopting $\epsilon_e$, we standardize the privacy settings across both methods for a meaningful comparison. 

Unless specified otherwise, we evaluate our approach and all comparative methods under $\epsilon_e$ values ranging from 0.01 to 10. For the GWAS outcomes, we consider the SNPs with a $p$-value less than $\alpha = 0.05$ as significant. After relaxation as mentioned in Section~\ref{sec:system}, the $p$-value threshold is increased by a tolerance factor $0.8$, resulting in a relaxed threshold of $\frac{\alpha}{0.8} = 0.0625$. All SNPs with $p$-values meeting this relaxed threshold in the reproduced outcomes are considered ``retained''. To evaluate robustness, we vary the error rate $\delta \in [0, 1]$ and generate $10$ copies of each dataset using our dataset sharing scheme. Each experiment is conducted as a single trial, and the average results are reported. In the graphical representations, shadowed regions indicate the $95\%$ confidence interval.

For the MIA evaluation against neural networks, we use a feedforward architecture designed for binary classification tasks. The network consists of four fully connected layers with sizes 512, 128, 32, and 1, respectively. We adopt LeakyReLU activation, consistent with its use in the GAN approach~\cite{yelmen2021creating}.

\subsection{GWAS Outcome Validation}
\label{sec:eval_gwas}

In this section, we evaluate the results of the GWAS outcome validation experiments. We first present the experimental results of the proposed scheme and the local differential privacy approach~\cite{kasiviswanathan2011can}, followed by a discussion on the performance of synthesis-based approaches. It is important to note that a higher retention rate difference at the same error rate indicates better performance, as it reflects greater detectability of errors in GWAS outcome validation.

\begin{figure*}[ht]
    \centering
    \includegraphics[width=\textwidth]{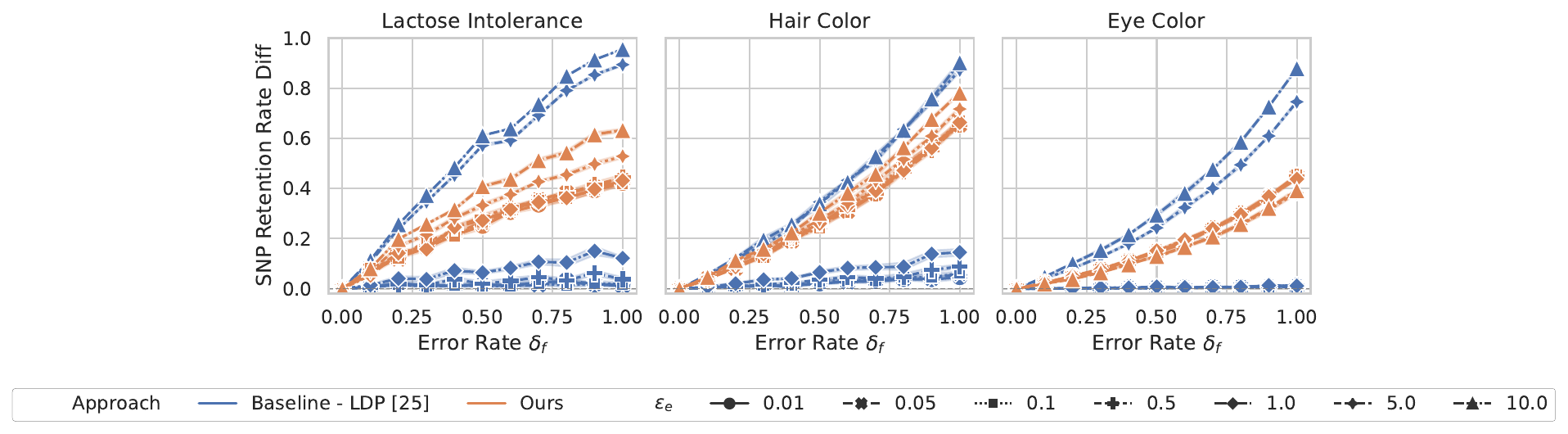}

    \caption{Performance of GWAS outcome validation for the $\chi^2$ test against flipping errors between ours and LDP~\cite{kasiviswanathan2011can}. 
    }
    \label{fig:chi2_flipping_error}
\end{figure*}
\begin{figure*}[ht]
    \centering
    \includegraphics[width=\textwidth]{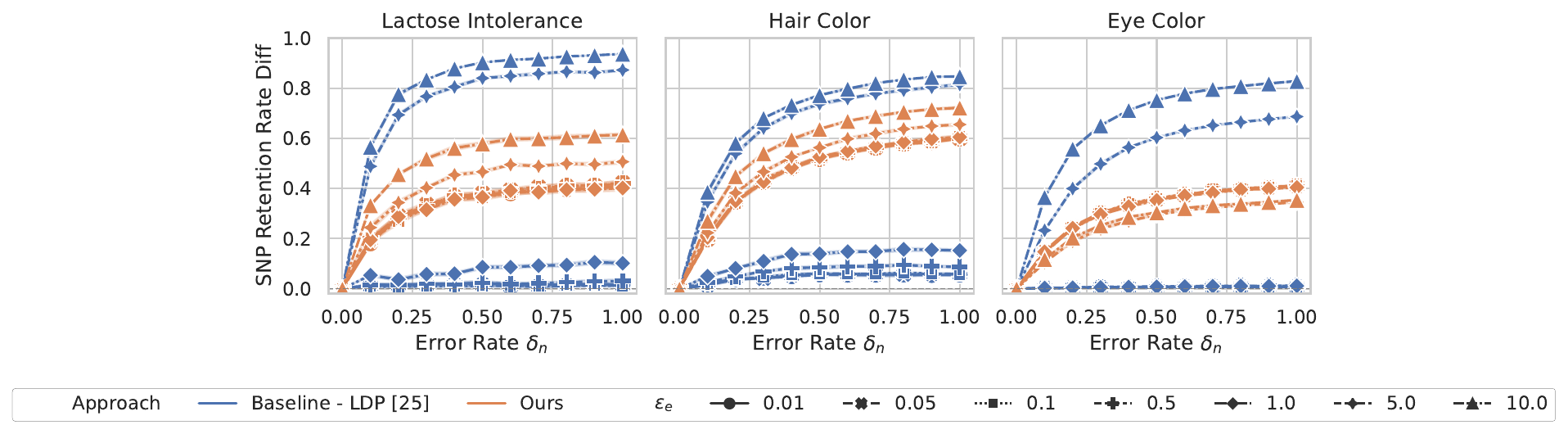}
    \caption{Performance of GWAS outcome validation for the $\chi^2$ test against noise errors between ours and LDP~\cite{kasiviswanathan2011can}. 
    }
    \label{fig:chi2_noise_error}
\end{figure*}

\begin{figure*}[ht]
    \centering
    \includegraphics[width=\textwidth]{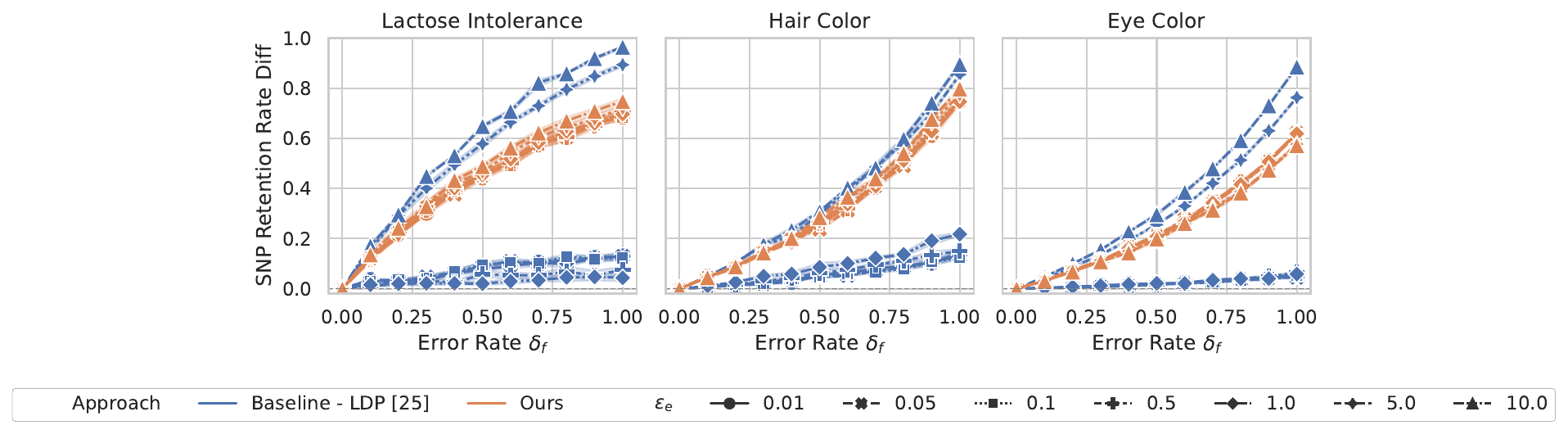}
    \caption{Performance of GWAS outcome validation for the odds ratio test against flipping errors between ours and LDP~\cite{kasiviswanathan2011can}. 
    }
    \label{fig:odd_flipping_error}
\end{figure*}

\begin{figure*}[ht]
    \centering
    \includegraphics[width=\textwidth]{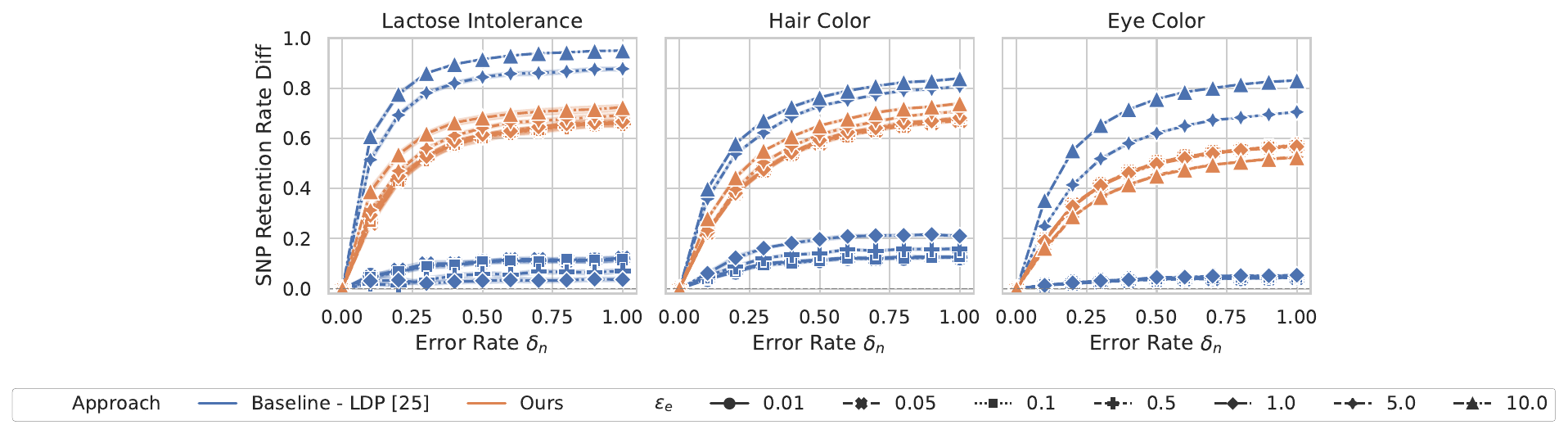}
    \caption{Performance of GWAS outcome validation for the odds ratio test against noise errors between ours and LDP~\cite{kasiviswanathan2011can}. 
    }
    \label{fig:odd_noise_error}

\end{figure*}

\subsubsection{$\chi^2$ Test}
\label{sec:eval_chi_utility}
Figure~\ref{fig:chi2_flipping_error} shows the experimental results for flipping errors in the $\chi^2$ test. Our method maintains a high SNP retention rate difference even with minor errors, demonstrating consistent performance across all three datasets. This strong detectability allows the verifier to reliably associate significant deviations with small errors. In contrast, LDP shows no noticeable difference when $\epsilon_e \leq 1$, making it ineffective in detecting even substantial errors. While LDP's performance improves significantly for $\epsilon_e \geq 5$, such a high privacy budget results in minimal data perturbation, making it susceptible to membership inference attacks (see Section~\ref{sec:eval_privacy}).

Figure~\ref{fig:chi2_noise_error} compares our scheme’s ability to detect noise errors against LDP. The trends are similar to those observed in Figure~\ref{fig:chi2_flipping_error}, though the curves exhibit sharper changes at lower error rates. This suggests that our approach is more sensitive to noise errors, providing stronger confidence in identifying such discrepancies. Overall, these results demonstrate that our method consistently outperforms LDP in detecting both flipping and noise errors, ensuring higher reliability in GWAS outcome validation.




\subsubsection{Odds Ratio Test}
\label{sec:eval_odd_utility}
In the odds ratio test, the retention rate differences observed in the presence of errors closely resemble those from the $\chi^2$ test, as shown in Figures~\ref{fig:odd_flipping_error} and~\ref{fig:odd_noise_error}. However, under acceptable privacy levels, the LDP approach consistently shows very low retention rate differences, making it ineffective for detecting errors in GWAS while maintaining privacy. These results further confirm that our method provides more reliable and accurate validation in GWAS compared to LDP.


\subsubsection{Comparison with Synthesis-based Approaches}
\label{sec:eval_gan_gwas}
We also implemented synthesis-based approaches, including the GAN approach~\cite{yelmen2021creating}, DPSyn~\cite{li2021dpsyn}, and PrivBayes~\cite{zhang2017privbayes}, for comparison. However, we encountered significant limitations during their implementation.

DPSyn and PrivBayes rely on internal correlations between features (SNPs in our case). As the dimensionality increases, the time cost grows non-linearly (as detailed in Section~\ref{sec:time}). This limits these methods to generating synthetic datasets only when the sample dimensionality is less than or equal to 140 SNPs. 

Meanwhile, the GAN approach, although able to complete the dataset generation, suffers from significant utility loss. For the lactose intolerance dataset, 51 out of 60 generated samples were identical, highlighting the ineffectiveness of the generation method. The results show a nearly zero difference across all datasets, indicating that the method fails to detect significant errors that occur during the GWAS process. Given this negligible difference, we do not present the figures, as they provide no meaningful insights.

Given these limitations, the synthesis-based approaches have proven impractical for GWAS outcome validation in real-world setting. In contrast, our proposed method offers a more robust and reliable solution, consistently outperforming both LDP and synthesis-based methods in detecting errors while maintaining privacy and utility.

    


\begin{table*}[ht]
\centering
\begin{tabular}{|c|c|cc|cc|cc|cc|}
\hline
                                     &              & \multicolumn{2}{c|}{Sample Error}      & \multicolumn{2}{c|}{Mean Error}        & \multicolumn{2}{c|}{Point Error}       & \multicolumn{2}{c|}{Variance Error}    \\
Dataset                              & $\epsilon_e$ & LDP~\cite{kasiviswanathan2011can} & Ours            & LDP~\cite{kasiviswanathan2011can} & Ours            & LDP~\cite{kasiviswanathan2011can} & Ours            & LDP~\cite{kasiviswanathan2011can} & Ours           \\ \hline
\multirow{7}{*}{Eye Color}           & 0.01         & 0.6200               & \textbf{0.3946} & 0.4630               & \textbf{0.0004} & 0.4429               & \textbf{0.3475} & 0.3895               & \textbf{0.0322} \\
                                     & 0.05         & 0.6116               & \textbf{0.3942} & 0.4568               & \textbf{0.0004} & 0.4369               & \textbf{0.3473} & 0.3875               & \textbf{0.0322} \\
                                     & 0.1          & 0.6010               & \textbf{0.3936} & 0.4488               & \textbf{0.0004} & 0.4293               & \textbf{0.3469} & 0.3848               & \textbf{0.0322} \\
                                     & 0.5          & 0.5114               & \textbf{0.3892} & 0.3820               & \textbf{0.0004} & 0.3654               & \textbf{0.3443} & 0.3556               & \textbf{0.0323} \\
                                     & 1            & 0.3955               & \textbf{0.3830} & 0.2955               & \textbf{0.0005} & \textbf{0.2825}      & 0.3405          & 0.3035               & \textbf{0.0324} \\
                                     & 5            & \textbf{0.0124}      & 0.3125          & 0.0098               & \textbf{0.0006} & \textbf{0.0088}      & 0.2904          & \textbf{0.0129}      & 0.0372          \\
                                     & 10           & \textbf{0.0000}      & 0.2214          & \textbf{0.0000}      & 0.0008          & \textbf{0.0000}      & 0.2131          & \textbf{0.0001}      & 0.0369          \\ \hline
\multirow{7}{*}{Hair Color}          & 0.01         & 0.6120               & \textbf{0.4133} & 0.4343               & \textbf{0.0041} & 0.4430               & \textbf{0.3647} & 0.3716               & \textbf{0.0396} \\
                                     & 0.05         & 0.6034               & \textbf{0.4121} & 0.4283               & \textbf{0.0041} & 0.4370               & \textbf{0.3639} & 0.3695               & \textbf{0.0396} \\
                                     & 0.1          & 0.5927               & \textbf{0.4109} & 0.4207               & \textbf{0.0041} & 0.4292               & \textbf{0.3633} & 0.3667               & \textbf{0.0397} \\
                                     & 0.5          & 0.5045               & \textbf{0.3994} & 0.3589               & \textbf{0.0040} & 0.3654               & \textbf{0.3561} & 0.3379               & \textbf{0.0396} \\
                                     & 1            & 0.3898               & \textbf{0.3827} & 0.2784               & \textbf{0.0041} & 0.2823               & \textbf{0.3452} & 0.2876               & \textbf{0.0392} \\
                                     & 5            & \textbf{0.0121}      & 0.2356          & 0.0111               & \textbf{0.0062} & \textbf{0.0088}      & 0.2271          & \textbf{0.0133}      & 0.0338          \\
                                     & 10           & \textbf{0.0000}      & 0.1411          & \textbf{0.0000}      & 0.0065          & \textbf{0.0000}      & 0.1381          & \textbf{0.0001}      & 0.0229          \\ \hline
\multirow{7}{*}{Lactose Intolerance} & 0.01         & 0.6153               & \textbf{0.4684} & 0.4366               & \textbf{0.0039} & 0.4429               & \textbf{0.4112} & 0.3276               & \textbf{0.0609} \\
                                     & 0.05         & 0.6072               & \textbf{0.4673} & 0.4309               & \textbf{0.0039} & 0.4369               & \textbf{0.4105} & 0.3261               & \textbf{0.0608}          \\
                                     & 0.1          & 0.5973               & \textbf{0.4654} & 0.4235               & \textbf{0.0038} & 0.4297               & \textbf{0.4093} & 0.3238               & \textbf{0.0610} \\
                                     & 0.5          & 0.5076               & \textbf{0.4536} & 0.3605               & \textbf{0.0038} & \textbf{0.3652}      & 0.4020          & 0.2993               & \textbf{0.0607} \\
                                     & 1            & \textbf{0.3926}      & 0.4353          & 0.2793               & \textbf{0.0041} & \textbf{0.2824}      & 0.39001         & 0.2557               & \textbf{0.0609} \\
                                     & 5            & \textbf{0.0123}      & 0.2532          & 0.0112               & \textbf{0.0072} & \textbf{0.0088}      & 0.2430          & \textbf{0.0127}      & 0.0496          \\
                                     & 10           & \textbf{0.0000}      & 0.1308          & \textbf{0.0000}      & 0.0082          & \textbf{0.0000}      & 0.1277          & \textbf{0.0000}      & 0.0279          \\ \hline
\end{tabular}

\caption{Data utility comparisons across datasets for our approach versus LDP~\cite{kasiviswanathan2011can}. Confidence intervals for all results are very small and thus omitted. Outcomes with superior results are highlighted in bold.}

\label{tab:utility}
\end{table*}

\subsection{Data Utility}
\label{sec:eval_utility}
Beyond evaluating GWAS reproducibility, we compared the performance of our scheme with LDP using the utility metrics outlined in Section~\ref{sec:utility_metrics}. The results, presented in Table~\ref{tab:utility}, show that our scheme outperforms LDP across all utility metrics for $\epsilon_e \leq 1$, except for point error in the eye color dataset and sample error in the lactose intolerance dataset at $\epsilon_e = 1$. When $\epsilon_e \geq 5$, LDP generally achieves better utility. However, this is because an LDP setting of $\epsilon_e \geq 5$ results in minimal data perturbation, preserving utility while significantly increasing vulnerability to membership inference attacks (MIAs), as shown in Figure~\ref{fig:attack_evaluation_eye_ml}. 

We further evaluate the utility of our method against PrivBayes~\cite{zhang2017privbayes} and DPSyn~\cite{li2021dpsyn} under a 100-SNP setting to assess performance on smaller datasets. This comparison is necessary because both PrivBayes and DPSyn are unable to handle larger-scale datasets, as discussed in Section~\ref{sec:eval_para}. To ensure a fair evaluation, we adopt a relaxed setting in which only 100 SNPs from each dataset are shared when comparing against these baselines. Due to space constraints, detailed results are provided in Appendix~\ref{sec:eval_additional_utililty}.

\subsection{Robustness Against Membership Inference Attacks}
\label{sec:eval_privacy}
We compared the robustness of our scheme against membership inference attacks (MIAs) with LDP across three datasets. Initially, we presented the comparison of the maximum attack power on these datasets in Table~\ref{tab:eval_max_mia}. A higher maximum attack power, defined as the highest value among all MIAs discussed in Section~\ref{eval:ml_metric}, signifies a greater risk of privacy breaches. Our scheme consistently outperformes LDP across all three datasets, with LDP exhibiting higher attack power in all scenarios.  

The maximum attack power on the hair color dataset reaches 0.75, originating from support vector machine attacks. As noted in Section~\ref{eval:ml_metric}, the datasets, particularly those for lactose intolerance and hair color, each contain only 60 individuals, which significantly limits their reliability against machine learning-based MIAs. Consequently, only the experiments conducted with the eye color dataset closely resemble real-world scenarios.  

\begin{figure*}
    \centering
    \includegraphics[width=\textwidth]{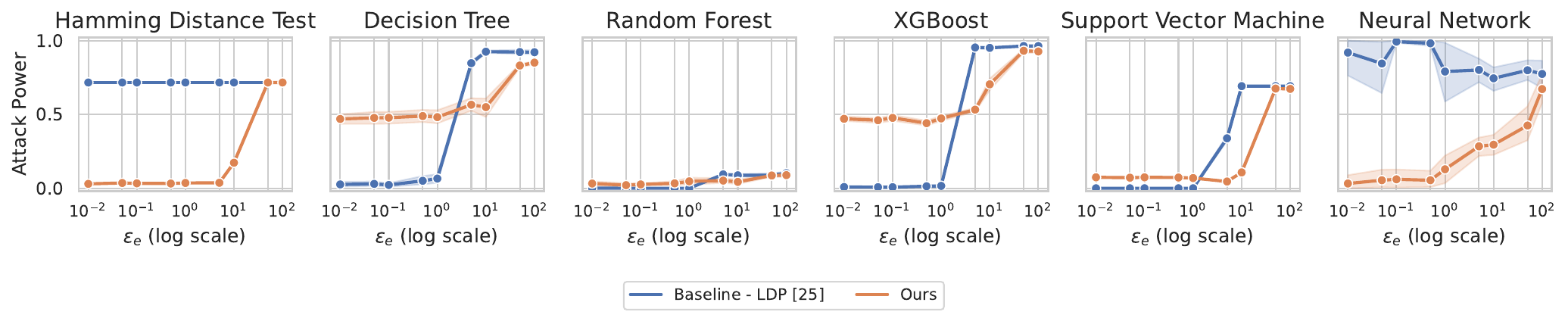}
    \caption{Comparison of our approach and local differential privacy (LDP)~\cite{kasiviswanathan2011can} against MIAs on the eye color dataset. Our scheme maintains low attack power for $\epsilon_e < 5$, while LDP remains vulnerable to Hamming distance and neural network attacks.}
    \label{fig:attack_evaluation_eye_ml}
\end{figure*}

The detailed experimental results for the eye color dataset are shown in Figure~\ref{fig:attack_evaluation_eye_ml}. We observe that LDP exhibits a consistently high attack success rate against the Hamming distance attack and the neural network attack, independent of $\epsilon_e$, indicating that it remains vulnerable regardless of the chosen privacy level. Additionally, both methods become susceptible to MIAs when $\epsilon_e$ reaches 5 or higher, primarily due to the effectiveness of decision trees and XGBoost in such attacks. However, at this point, the perturbation is minimal, making these cases unrealistic for practical consideration.  

Under acceptable privacy levels, our scheme consistently outperforms LDP in mitigating these vulnerabilities. Due to space constraints, the detailed experimental results for the remaining two datasets are provided in Appendix~\ref{sec:addition_exp_ml}.

\begin{table*}[ht]
\centering
\footnotesize
\resizebox{0.8\textwidth}{!}{
\begin{tabular}{llccccccc}
\toprule
\multicolumn{1}{c}{}                 & \multicolumn{1}{c}{}    & \multicolumn{7}{c}{$\epsilon_e$}                                                                                     \\ \cline{3-9} 
Dataset                              & Approach                & 0.01           & 0.05           & 0.1            & 0.5            & 1              & 5              & 10             \\ \hline
\multirow{2}{*}{Eye Color}           & Baseline - LDP~\cite{kasiviswanathan2011can} & 0.967          & 0.93           & 0.993          & 0.981          & 0.91           & 0.953          & 0.952          \\
                                     & Ours                    & \textbf{0.489} & \textbf{0.499} & \textbf{0.511} & \textbf{0.499} & \textbf{0.514} & \textbf{0.574} & \textbf{0.679} \\ \hline
\multirow{2}{*}{Hair Color}          & Baseline - LDP~\cite{kasiviswanathan2011can} & 0.833          & 0.833          & 0.833          & 0.833          & 0.833          & 0.883          & 0.917          \\
                                     & Ours                    & \textbf{0.75}  & \textbf{0.75}  & \textbf{0.75}  & \textbf{0.75}  & \textbf{0.75}  & \textbf{0.833} & \textbf{0.833} \\ \hline
\multirow{2}{*}{Lactose Intolerance} & Baseline - LDP~\cite{kasiviswanathan2011can} & 0.883          & 0.875          & 0.892          & 0.883          & 0.867          & 0.842          & 0.833          \\
                                     & Ours                  & \textbf{0.558} & \textbf{0.558} & \textbf{0.525} & \textbf{0.517} & \textbf{0.533} & \textbf{0.75}  & \textbf{0.858} \\ 

\bottomrule
\end{tabular}
}

\caption{Comparison of maximum attack power against MIAs. Better results are marked in bold.}

\label{tab:eval_max_mia}

\end{table*}

\subsection{Time Complexity}
\label{sec:time}

\begin{figure}[ht]
 \centering
  \includegraphics[width=0.4\textwidth]{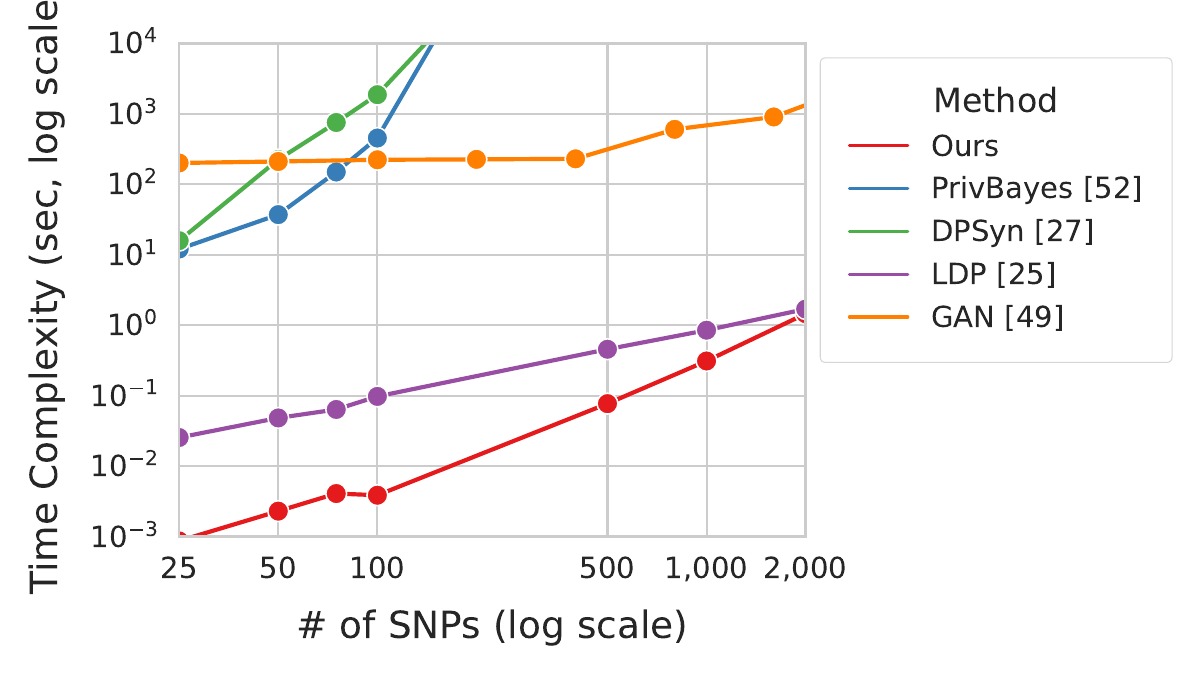}

  \caption{Time complexity.}

\label{fig:time}

\end{figure}

We compare the time complexity among our scheme, local differential privacy (LDP)~\cite{kasiviswanathan2011can}, DPSyn~\cite{li2021dpsyn}, PrivBayes~\cite{zhang2017privbayes}, and GAN~\cite{yelmen2021creating}, on the lactose intolerance dataset using varied number of SNPs in the dataset to demostrate the time consumption of the generation for each metchod. Figure~\ref{fig:time} displays the results. Both our scheme and LDP outperform the synthesis-based approaches, with our scheme exhibiting slightly lower time complexity when the number of SNPs is below 2,000. Beyond this threshold, LDP demonstrates better performance; however, as discussed in Section~\ref{sec:eval_privacy}, its vulnerability to membership inference attacks (MIA) makes it unsuitable for secure data sharing.  

The two synthetic methods, namely DPSyn and PrivBayes, require substantial time for data generation as the number of SNPs increases. Specifically, their time cost escalates sharply, reaching $10^5$ seconds when the SNP count exceeds 150. These findings highlight that our scheme is significantly more time-efficient than synthesis-based methods, making it a more practical solution for large-scale genomic data sharing.

\section{Limitations and Future Work}
\label{sec:conclusion}

In this paper, we propose a novel scheme for sharing genomic datasets in a privacy-preserving manner, specifically for GWAS outcome validation. We efficiently adapted the XOR mechanism to generate binary datasets while preserving correlations with the help of published Minor Allele Frequencies (MAFs). Our approach demonstrates superiority in detecting GWAS outcome errors, maintaining data utility, and providing robustness against membership inference attacks.  

While our approach is specifically optimized for enhancing GWAS reproducibility, it has certain limitations. It may not generalize well to other genomic studies, such as transcriptome-wide association studies, genetic epidemiology, or gene-environment interaction analyses. Additionally, our method does not explicitly address scenarios where malicious researchers fabricate datasets to report false results. However, the likelihood of such misconduct is low, as ethical risks and potential career repercussions serve as strong deterrents.  

Despite these limitations, our method remains highly effective within its intended scope, significantly improving the reproducibility and utility of GWAS outcomes. Moving forward, we will explore strategies to further optimize privacy and usability in practical genomic research settings. Future work will also focus on integrating dataset fingerprinting techniques to enhance accountability and strengthen privacy assurances in genomic data sharing.

\bibliographystyle{plain}
\bibliography{bib.bib}

\appendix
\label{sec:appendix}

\section{Symbols and Notations}
\label{sec:symbols}
We show frequent symbols and notations used in the paper in Table~\ref{tab:symbol}.

\begin{table}[h]

\footnotesize
  \centering

\begin{tabular}{|p{1cm}|p{6cm}| }
 \hline
 \textbf{Notations} & \textbf{Descriptions} \\
 \hline
$n$ & The number of individuals in the dataset $D$ \\
$m$ & The number of SNPs in the dataset $D$ \\
$D^b$ & The binarized version of $D$ \\
$\tilde{D}^b$ & The perturbed (binarized) dataset from Stage 1 \\
$\hat{D}^b$ & The utility-restored (binarized) dataset from Stage 2 \\
$D'$ & The output dataset \\
$\epsilon_e$ & Effective privacy budget for each SNP \\
 \hline
\end{tabular}
\caption{\label{tab:symbol}Symbols and notations.}
\end{table}

\section{Proof of Theorem~\ref{thm:relaxed-xor}}
\label{app:proof-relaxed-xor}

\begin{proof} In the first step, we denote
\begin{equation*}
    \resizebox{0.9\hsize}{!}{$\mathcal{S}_u =\big\{\bb|\bb_u=1, \bb_v\in\{0,1\},\forall v\in\{1,2,\cdots,NP\}, v\neq u\big\}.$}
\end{equation*}
Then, we bound the marginal probability of each noise bit taking value 1 as follows 
\begin{equation*}
    \begin{aligned}
 \Pr[\mathbf{b}_u = 1] & =    \sum_{\bb\in\mathcal{S}_u} C(\bm{\Pi})\exp\{\bb^T\bm{\Pi}\bb\}  \\  
 &= \frac{\sum_{\bb\in\mathcal{S}_u} \exp\{\bb^T\bm{\Pi}\bb\}}{\sum_{\bb\in\mathcal{S}} \exp\{\bb^T\bm{\Pi}\bb\}} \\  
 &\stackrel{(a)}= \resizebox{0.7\hsize}{!}{$\frac{\sum_{\bb\in\mathcal{S}_u} \exp\{\bb^T\bm{\Pi}\bb\}}{\sum_{\bb\in\mathcal{S}_u} \exp\{\bb^T\bm{\Pi}\bb\}+\sum_{\bb\in\overline{\mathcal{S}_u}} \exp\{\bb^T\bm{\Pi}\bb\}}$}\\  
 &= \frac{1}{1+\frac{\sum_{\bb\in\overline{\mathcal{S}_u}} \exp\{\bb^T\bm{\Pi}\bb\}}{\sum_{\bb\in\mathcal{S}_u} \exp\{\bb^T\bm{\Pi}\bb\}}}\\ 
 &\stackrel{(b)}= \frac{1}{1+\frac{\sum_{\bb\in\overline{\mathcal{S}_u}} \exp\{\bb^T\bm{\Pi}\bb\}}{\sum_{\bb\in\overline{\mathcal{S}_u}} \exp\left\{(\bb+\boldsymbol{j}_u)^T\bm{\Pi}(\bb+\boldsymbol{j}_u)\right\}  }} \\  
 &\stackrel{(c)}\leq \frac{1}{1+\min_{\bb\in\overline{\mathcal{S}_u}}\frac{\exp\{\bb^T\bm{\Pi}\bb\}}{ \exp\left\{(\bb+\boldsymbol{j}_u)^T\bm{\Pi}(\bb+\boldsymbol{j}_u)\right\}  }}\\  
 &\leq \frac{1}{1+\max_{\bb\in\overline{\mathcal{S}_u}}  \exp\left\{ 2\boldsymbol{j}_u^T\bm{\Pi}\bb+\bm{\Pi}_{u,u} \right \}  },
 \end{aligned}
 \end{equation*}
where in $(a)$, $\overline{\mathcal{S}_u}$ is the complementary set of $\mathcal{S}_u$, i.e.,  
\begin{equation*}
    \resizebox{0.9\hsize}{!}{$\overline{\mathcal{S}_u} =\big\{\bb|\bb_u=0, \bb_v\in\{0,1\},\forall v\in\{1,2,\cdots,NP\}, v\neq u\big\}.$}
\end{equation*}
$(b)$ is because by defining  the one-hot vector $$\boldsymbol{j}_u\in\{0,1\}^{NP\times 1}$$ that only has 1 at the $u$th position, and 0 at all the other positions. Then,  $\forall \bb\in\overline{\mathcal{S}_u} $, we have $\bb+\boldsymbol{j}_u\in\mathcal{S}_u$. $(c)$ is because
$\frac{\sum x_i}{\sum y_i}\geq \min_i \frac{x_i}{y_i}$ for positive sequences $x_i$ and $y_i$, and in $(c)$ $\bm{\Pi}_{u,u}$ represents the entry of $\bm{\Pi}$ in the $u$th row and $u$th column.

In step 2, we proceed to calculate the maximum value, i.e., 
\begin{equation*}
  \resizebox{0.99\hsize}{!}{$  \max_{\bb\in\overline{\mathcal{S}_u}}  \exp\left\{ 2\boldsymbol{j}_u^T\bm{\Pi}\bb+\bm{\Pi}_{u,u} \right \}  =  \exp\left\{ \bm{\Pi}_{u,u}+ \max_{\bb\in\overline{\mathcal{S}_u}} 2\boldsymbol{j}_u^T\bm{\Pi}\bb  \right \}.$}
\end{equation*}
 In particular, we observe that $\boldsymbol{j}_u^T\bm{\Pi}$ represents the $u$th row of $\bm{\Pi}$, thus $\max_{\bb\in\overline{\mathcal{S}_u}} \boldsymbol{j}_u^T\bm{\Pi}\bb$ corresponds to the summation of all positive values in the $u$th row of $\bm{\Pi}$ except for $\bm{\Pi}_{u,u}$ (since $\bb_u = 0$). 
We denote $\kappa_u = 2\times  {\mathrm Sum}(\bm{\Pi}_u) - \bm{\Pi}_{u,u}$.

In step 3, we prove that the probability ratio of the outputs of the efficient genomic dataset perturbation is bounded by $\exp(\epsilon)$. W.l.o.g., suppose $D$ and $D'$ only differ by the SNP sequence of the first  individual, and let $\boldsymbol{d}$ and $\boldsymbol{d}'$ be the encoded SNP sequences of the first individual in $D$ and $D'$, respectively.

\begin{gather*}
    \begin{aligned}
       & \ln\left( \frac{\prod_u \Pr(\boldsymbol{d}_u\oplus B_u = O_u)}{\prod_u \Pr(\boldsymbol{d}'_u\oplus B'_u = O_u)} \right)\\ 
= & \sum_u\ln \frac{ \Pr( B_u = O_u \oplus \boldsymbol{d}_u)}{ \Pr( B'_u = O_u \oplus \boldsymbol{d}'_u )}\\ 
= & \sum_u\ln\frac{ \Pr( B_u = 1 )^{O_u \oplus \boldsymbol{d}_u}(1-\Pr( B_u = 1 ))^{1-(O_u \oplus \boldsymbol{d}_u)}}{ \Pr( B_u' = 1 )^{O_u \oplus \boldsymbol{d}_u'}(1-\Pr( B_u' = 1 ))^{1-(O_u \oplus \boldsymbol{d}_u')}}\\ 
=&\sum_u [(O_u \oplus \boldsymbol{d}_u)-(O_u \oplus \boldsymbol{d}_u')] \ln \Pr(B_u=1) +\\
&\quad\ \ [(O_u \oplus \boldsymbol{d}_u')-(O_u \oplus \boldsymbol{d}_u)] \ln (1-\Pr(B_u=1))\\
=&\Biggl(\sum_{\{u:\kappa_u>||\lambda(\T)||\}} [(O_u \oplus \boldsymbol{d}_u)-(O_u \oplus \boldsymbol{d}_u')] \ln \Pr(B_u=1) +\\
&\quad\ \ [(O_u \oplus \boldsymbol{d}_u')-(O_u \oplus \boldsymbol{d}_u)] \ln (1-\Pr(B_u=1)) \Biggl)  \\
+ &\Biggl(\sum_{\{u:\kappa_u\leq||\lambda(\T)||\}} [(O_u \oplus \boldsymbol{d}_u)-(O_u \oplus \boldsymbol{d}_u')] \ln \Pr(B_u=1) +\\
&\quad\ \ [(O_u \oplus \boldsymbol{d}_u')-(O_u \oplus \boldsymbol{d}_u)] \ln (1-\Pr(B_u=1)) \Biggl)\\
\stackrel{(a)}= &\sum_{\{u:\kappa_u\leq||\lambda(\T)||\}} [(O_u \oplus \boldsymbol{d}_u)-(O_u \oplus \boldsymbol{d}_u')] \ln \frac{1}{1+\exp(\kappa_u)} +\\
&\quad\ \ [(O_u \oplus \boldsymbol{d}_u')-(O_u \oplus \boldsymbol{d}_u)] \ln \frac{\exp(\kappa_u)}{1+\exp(\kappa_u)}\\
=&\sum_{\{u:\kappa_u\leq||\lambda(\T)||\}}  [(O_u \oplus \boldsymbol{d}_u)-(O_u \oplus \boldsymbol{d}_u')] \\
& \qquad\quad \times  \left( \ln \frac{1}{1+\exp(\kappa_u)}-\ln \frac{\exp(\kappa_u)}{1+\exp(\kappa_u)}\right)\\
=&\sum_{\{u:\kappa_u\leq||\lambda(\T)||\}}  [(O_u \oplus \boldsymbol{d}_u)-(O_u \oplus \boldsymbol{d}_u')]   \ln \frac{1}{\exp(\kappa_u)}\\
\stackrel{(b)}=&\sum_{\{u:\kappa_u\leq||\lambda(\T)||\}}    |2O_u-1| |\boldsymbol{d}_u'-\boldsymbol{d}_u| |\kappa_u|\\
\stackrel{(c)}< & s_f ||\lambda(\bm{\Pi})||_2 \leq s_f\big(||\boldsymbol{\lambda}(\T)||_2 +   \textstyle\sum_{i=1}^{n-1}\sum_{j= i+1}^n  ||\boldsymbol{\lambda}(\La_{i,j})||_2 \big) ,
    \end{aligned}
\end{gather*}
where $(a)$ is because the summation is 0 for $\{u:\kappa_u>||\lambda(\T)||\}$, $(b)$ is because $u\oplus v = (1-u)v+u(1-v)$ for binary $u$ and $v$, and $(c)$ is because the cardinity of set $\{u:\kappa_u\leq||\lambda(\T)||\}$ is at most $s_f$. According to (\ref{condition_general_T}), we can  complete the proof. 
\end{proof}

\section{Further Details About the Post-Processing}
\label{sec:detail_post_processing}

We use the  Minor Allele Frequencies (MAFs) published in the research findings, denoted as $\mathcal{M}^r$, as a reference and calculate the MAFs in the noisy dataset $\tilde{D}^b$ as $\tilde{\mathcal{M}}$. We first convert these MAFs into binary distributions with percentages of 0's and 1's, denoted as $\mathcal{C}_j$ for $\mathcal{M}^r_j$ and $\tilde{\mathcal{C}}_j$ for $\tilde{\mathcal{M}}_j$. Our goal is to adjust the distribution of 0's and 1's in $\tilde{D}^b$ so that the final dataset, denoted as $\hat{D}^b$, aligns its MAFs closely with the public MAFs.

First, we define the cost as $\mathbb{C}_{pq}=\vert p - q\vert$ for the change from $p$ to $q$, where $p, q \in \{0, 1\}$, aiming to modify the dataset with the minimum total cost.
We then construct an optimization problem at each position $j$ in the encoded genomic dataset $\tilde{D}^b$. We consider $C'_{j}$ and $\tilde{C}_{j}$ as two mass distributions at position $j$, aiming to find a transport plan $T^{\mathcal{K}\times \mathcal{K}}$ that modifies the mass of $C'_{j}$ to make it resemble $\tilde{C}_{j}$. The total cost is defined as:
\begin{equation}
    \langle T, \mathbb{C} \rangle = \sum_{p=0}^{1} \sum_{q=0}^{1} T_{pq}\mathbb{C}_{pq},
\end{equation}
where $T$ is the transport plan and $\mathbb{C}$ is the matrix of costs.

The optimal transport is formulated as follows:
\begin{align*}
    \min_T & \langle T, \mathbb{C} \rangle \\
    \text{s.t. } & \sum_{q=0}^{1} T_{pq} = c'^{\text{norm}}_{pj}\,\, \forall p \in \{0, 1\} \\
                 & \sum_{p=0}^{1} T_{pq} = \tilde{c}^{\text{norm}}_{qj}\,\, \forall q \in \{0, 1\} \\
                 & T_{pq} \geq 0 \,\, \forall (p,q) \in \{0, 1\} \times \{0, 1\},
\end{align*}
where $\tilde{C}^{\text{norm}}_{kj}$ and $C'^{\text{norm}}_{kj}$ are normalized as:
\begin{equation}
    \tilde{C}^{\text{norm}}_{kj} = \frac{\tilde{c}_{kj}}{|\tilde{C}_{j}|}, \, C'^{\text{norm}}_{kj} = \frac{c'_{kj}}{|C'_{j}|}, \, k \in \{0, 1\}.
\end{equation}
Here, $\tilde{c}_{kj}$ and $c'_{kj}$ are the counts of SNP value $k$ at position $j$ in $\tilde{\mathcal{C}}$ and $\mathcal{C}'$, respectively. $T$ essentially represents a joint mass distribution at each position $j$ ($\sum_p \sum_q T_{pq} = 1, \forall j$) whose row- or column-wise marginalization is the marginal distribution of SNP taking value $p$ or $q$ at position $j$.

This one-dimensional optimization problem is solved using optimal transport (OT), a method in transportation theory aimed at minimizing the cost while transferring the distribution from one state to another. We use the existing Python package \cite{flamary2021pot} to calculate this one-dimensional optimal transport, applying the formulated strategy to adjust SNP values based on $T_{pq}$, effectively transferring $\left\lfloor T_{pq} \times n \right\rfloor$ alleles from one category to another.

\section{Details of the Experiment Results Against MIAs}
\label{sec:addition_exp_ml}
\begin{figure*}
    \centering
    \includegraphics[width=\textwidth]{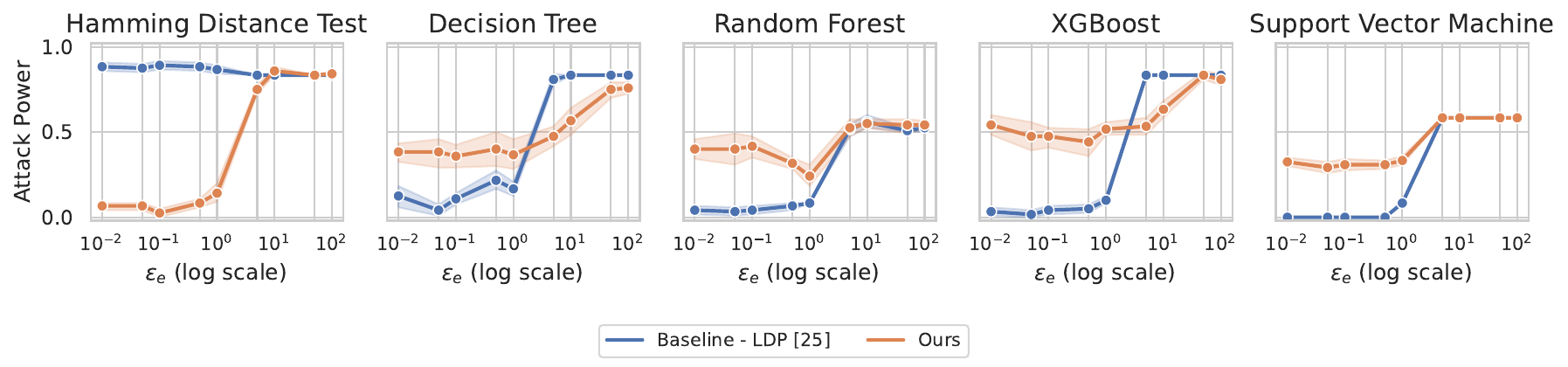}
    \caption{Comparison of robustness of our approach and local differential privacy (LDP)~\cite{kasiviswanathan2011can} against different membership inference attacks on the lactose intolerance dataset.}
    \label{fig:attack_evaluation_lactose_ml}
\end{figure*}

\begin{figure*}
    \centering
    \includegraphics[width=\textwidth]{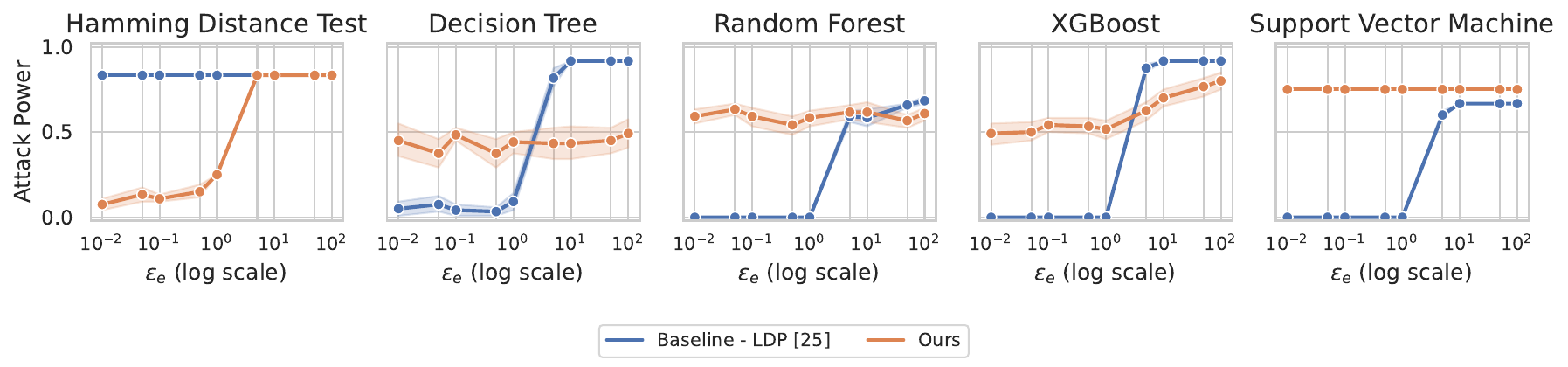}
    \caption{Comparison of robustness of our approach and local differential privacy (LDP)~\cite{kasiviswanathan2011can} against different membership inference attacks on the hair color dataset.}
    \label{fig:attack_evaluation_hair_ml}
\end{figure*}

We present detailed experimental results on the lactose intolerance and hair color datasets against MIAs in Figures~\ref{fig:attack_evaluation_lactose_ml} and~\ref{fig:attack_evaluation_hair_ml}.

\section{Additional Utility Comparison}
\label{sec:eval_additional_utililty}
In our analysis, we conduct a comprehensive utility comparison of our method against DPSyn~\cite{li2021dpsyn} and PrivBayes~\cite{zhang2017privbayes} across three toy datasets, each comprising 100 SNPs, as shown in Figure~\ref{tab:utility-100}.

\begin{table*}
  \centering
  \footnotesize
  \begin{sideways}
\begin{tabular}{|c|c|ccc|ccc|ccc|ccc|}
\hline
                                     & Utility Metric & \multicolumn{3}{c|}{Sample Error}    & \multicolumn{3}{c|}{Mean Error}      & \multicolumn{3}{c|}{Point Error}     & \multicolumn{3}{c|}{Variance Error}        \\
                                     & Approach       & DPSyn  & PrivBayes & Ours            & DPSyn  & PrivBayes & Ours            & DPSyn  & PrivBayes & Ours            & DPSyn  & PrivBayes       & Ours            \\
Dataset                              & $\epsilon_e$   &        &           &                 &        &           &                 &        &           &                 &        &                 &                 \\ \hline
\multirow{7}{*}{Lactose Intolerance} & 0.01           & 0.8818 & 0.8766    & \textbf{0.4654} & 0.6107 & 0.6071    & \textbf{0.0041} & 0.6476 & 0.6454    & \textbf{0.4075} & 0.2973 & 0.2619          & \textbf{0.0604} \\
                                     & 0.05           & 0.8818 & 0.8514    & \textbf{0.4717} & 0.6107 & 0.5650    & \textbf{0.0040} & 0.6476 & 0.6276    & \textbf{0.4132} & 0.2973 & 0.2662          & \textbf{0.0611} \\
                                     & 0.1            & 0.8818 & 0.8216    & \textbf{0.4663} & 0.6107 & 0.5201    & \textbf{0.0038} & 0.6476 & 0.6096    & \textbf{0.4094} & 0.2973 & 0.2672          & \textbf{0.0622} \\
                                     & 0.5            & 0.8818 & 0.6890    & \textbf{0.4550} & 0.6107 & 0.3292    & \textbf{0.0035} & 0.6476 & 0.5279    & \textbf{0.4033} & 0.2973 & 0.2406          & \textbf{0.0603} \\
                                     & 1              & 0.8818 & 0.6052    & \textbf{0.4362} & 0.6107 & 0.2039    & \textbf{0.0042} & 0.6476 & 0.4775    & \textbf{0.3897} & 0.2973 & 0.1750          & \textbf{0.0608} \\
                                     & 5              & 0.8818 & 0.5085    & \textbf{0.2585} & 0.6107 & 0.0789    & \textbf{0.0072} & 0.6476 & 0.4184    & \textbf{0.2474} & 0.2973 & 0.0807          & \textbf{0.0508} \\
                                     & 10             & 0.8818 & 0.5050    & \textbf{0.1365} & 0.6107 & 0.0682    & \textbf{0.0085} & 0.6476 & 0.4182    & \textbf{0.1332} & 0.2973 & 0.0680          & \textbf{0.0300} \\ \hline
\multirow{7}{*}{Hair Color}          & 0.01           & 0.6986 & 0.7021    & \textbf{0.4199} & 0.4193 & 0.4539    & \textbf{0.0039} & 0.5517 & 0.5499    & \textbf{0.3718} & 0.2233 & 0.1882          & \textbf{0.0395} \\
                                     & 0.05           & 0.6986 & 0.6702    & \textbf{0.4188} & 0.4193 & 0.4146    & \textbf{0.0043} & 0.5517 & 0.5280    & \textbf{0.3717} & 0.2233 & 0.1863          & \textbf{0.0410} \\
                                     & 0.1            & 0.6986 & 0.6438    & \textbf{0.4174} & 0.4193 & 0.3749    & \textbf{0.0043} & 0.5517 & 0.5141    & \textbf{0.3701} & 0.2233 & 0.1836          & \textbf{0.0407} \\
                                     & 0.5            & 0.6986 & 0.5386    & \textbf{0.4054} & 0.4193 & 0.2109    & \textbf{0.0040} & 0.5517 & 0.4433    & \textbf{0.3634} & 0.2233 & 0.1487          & \textbf{0.0390} \\
                                     & 1              & 0.6986 & 0.4842    & \textbf{0.3872} & 0.4193 & 0.1292    & \textbf{0.0039} & 0.5517 & 0.4075    & \textbf{0.3507} & 0.2233 & 0.1053          & \textbf{0.0401} \\
                                     & 5              & 0.6986 & 0.4405    & \textbf{0.2396} & 0.4193 & 0.0610    & \textbf{0.0067} & 0.5517 & 0.3811    & \textbf{0.2314} & 0.2233 & 0.0514          & \textbf{0.0321} \\
                                     & 10             & 0.6986 & 0.4349    & \textbf{0.1411} & 0.4193 & 0.0566    & \textbf{0.0065} & 0.5517 & 0.3770    & \textbf{0.1384} & 0.2233 & 0.0490          & \textbf{0.0220} \\ \hline
\multirow{7}{*}{Eye Color}           & 0.01           & 0.8475 & 0.8075    & \textbf{0.3911} & 0.6137 & 0.5717    & \textbf{0.0004} & 0.6239 & 0.5952    & \textbf{0.3448} & 0.3435 & 0.2989          & \textbf{0.0327} \\
                                     & 0.05           & 0.8475 & 0.6634    & \textbf{0.3900} & 0.6137 & 0.3778    & \textbf{0.0004} & 0.6239 & 0.5044    & \textbf{0.3441} & 0.3435 & 0.2829          & \textbf{0.0324} \\
                                     & 0.1            & 0.8475 & 0.5603    & \textbf{0.3908} & 0.6137 & 0.2327    & \textbf{0.0005} & 0.6239 & 0.4406    & \textbf{0.3444} & 0.3435 & 0.2163          & \textbf{0.0322} \\
                                     & 0.5            & 0.8475 & 0.4412    & \textbf{0.3859} & 0.6137 & 0.0666    & \textbf{0.0005} & 0.6239 & 0.3662    & \textbf{0.3417} & 0.3435 & 0.0791          & \textbf{0.0319} \\
                                     & 1              & 0.8475 & 0.4169    & \textbf{0.3795} & 0.6137 & 0.0368    & \textbf{0.0005} & 0.6239 & 0.3522    & \textbf{0.3382} & 0.3435 & 0.0428          & \textbf{0.0333} \\
                                     & 5              & 0.8475 & 0.4034    & \textbf{0.3119} & 0.6137 & 0.0230    & \textbf{0.0007} & 0.6239 & 0.3443    & \textbf{0.2902} & 0.3435 & \textbf{0.0222} & 0.0377          \\
                                     & 10             & 0.8475 & 0.4027    & \textbf{0.2260} & 0.6137 & 0.0236    & \textbf{0.0007} & 0.6239 & 0.3445    & \textbf{0.2173} & 0.3435 & \textbf{0.0220} & 0.0380          \\ \hline
\end{tabular}

  \end{sideways}    
  \caption{Comprehensive comparison in data utility across three 100-SNP toy datasets for our approach versus DPSyn~\cite{li2021dpsyn} and PrivBayes~\cite{zhang2017privbayes}. Outcomes with superior results are highlighted in bold.}
    \label{tab:utility-100}
\end{table*}

\end{document}